\documentclass[a4paper,fleqn,usenatbib]{mnras}
\usepackage[T2A, T1]{fontenc}
\usepackage{ae,aecompl}
\usepackage{graphicx}
\usepackage{amsmath}
\usepackage{amssymb}
\usepackage{array,tabularx}

\usepackage[english]{babel}
\usepackage{threeparttable}
\usepackage{enumitem}
\title[KV UMa: BH mass and non-stellar spectrum]{Optical, \textit{J}, and \textit{K} light curves of XTE~J1118+480~=~KV~UMa: the mass of the black hole and the spectrum of the non-stellar component}

\author[A. M. Cherepashchuk et al.]{A. M. Cherepashchuk$^{1}$,\thanks{E-mail:Cherepashchuk@gmail.com (AMC)}
N. A. Katysheva$^{1}$,
T. S. Khruzina$^{1}$,
S. Yu. Shugarov$^{1,2}$,
\newauthor
A. M. Tatarnikov$^{1}$,
and A. I. Bogomazov$^{1}$
\\
$^{1}$M. V. Lomonosov Moscow State University, P. K. Sternberg Astronomical Institute, 119991, Universitetkij prospect, 13, Moscow, Russia\\
$^{2}$Astronomical Institute, Slovak Academy of Sciences, 05960, Tatransk\'a Lomnica, Slovakia\\
}

\date{Accepted 2019 September 11. Received 2019 August 17; in original form 2019 June 10}

\pubyear{2019}

\begin{document}
\label{firstpage}
\pagerange{\pageref{firstpage}--\pageref{lastpage}}
\maketitle

\begin{abstract}
Optical, \textit{J}, and \textit{K} photometric observations of the KV UMa black hole X-ray nova in its quiescent state obtained in 2017-2018 are presented. A significant flickering within light curves was not detected, although the average brightness of the system faded by $\approx 0.1^m$ during 350 days. Changes in the average brightness were not accompanied with the increase or the decrease of the flickering. From the modelling of five light curves the inclination of the KV UMa orbit and the black hole mass were obtained: $i=74^{\circ}\pm 4^{\circ}$, $M_{BH}=(7.06\div 7.24)M_{\odot}$ dependently on the used mass ratio. The non-stellar component of the spectrum in the range $\lambda=6400\div 22000$\r{A} can be fitted by a power law $F_{\lambda}\sim \lambda^{\alpha}$, $\alpha\approx -1.8$. The accretion disk orientation angle changed from one epoch to another. The model with spots on the star was inadequate. Evolutionary calculations using the ``Scenario Machine'' code were performed for low mass X-ray binaries, a recently discovered anomalously rapid decrease of the orbital period was taken into account. We showed that the observed decrease can be consistent with the magnetic stellar wind of the optical companion which magnetic field was increased during the common envelope stage. Several constraints on evolutionary scenario parameters were done.
\end{abstract}


\section{Inroduction}

X-ray novae are the main source of information about masses of black holes in X-ray binary systems (see, e.g., \citealp{casares2014} and their references). They are low mass X-ray binaries, where the low mass K-M optical star fills its Roche lobe and its matter outflows onto the relativistic object (a neutron star or a black hole). An accretion disk forms around the compact star. Most of the time X-ray novae are in quiescence, when their X-ray luminosity is small ($\lesssim 10^{31}\div 10^{33}$ erg s$^{-1}$). Instabilities in the accretion disk lead to the increase of the disk's matter turbulence and to the increase of the accretion rate. It leads to the outburst of the X-ray radiation with the duration about a month. The maximum X-ray luminosity during the outburst is about $10^{36}\div 10^{38}$ erg s$^{-1}$. The X-ray outburst is accompanied with the optical outburst due to the heating of the disk and the star by the powerful X-ray radiation. In quiescence the system's spectrum contains absorption lines of the optical star. The main cause of the optical variability of X-ray novae in quiescence is the ellipticity effect \citep{lyutyi1973}. The optical radiation of the accretion disk and of the region of interaction between the gas stream and the disk also gives an important contribution to the system's optical luminosity in low mass X-ray binaries. A significant part of X-ray novae contains a black hole as a relativistic object.

Recently it was realised that even in quiescence there are non-stationary processes in X-ray novae with black holes. They are manifested in the fact that some X-ray novae in quiescence show passive and active states of the optical variability \citep{cantrell2008,cantrell2010,cherepashchuk2019}. In the passive state the amplitude of the irregular variability (flickering) is relatively low, the orbital light curve has a regular shape. In the active state the average brightness of the system grows by several tenths of stellar magnitude, the amplitude of the flickering sharply grows. The orbital light curve at this state has irregular changes.

Besides, in black hole X-ray novae A0620-00 and Swift 71354.2-0933 the linear polarization of the IR radiation was found, it can indicate the synchrotron radiation of relativistic jets \citep{russell2016}. Also in these systems the anomalously fast decrease of the orbital period was found \citep{g-h-2012,g-h-2014,g-h-2017}, spectroscopic effects of the precession of the elliptical accretion disk \citep{shahbaz2004,zurita2016} were detected, and there were observed spectroscopic traces of chromospheric activity of the donor star \citep{casares1997,torres2002,zurita2003,g-h-2010,zurita2016}.

X-ray binary XTE J1118+480 = KV UMa belongs to a class of low mass transient X-ray binary systems with black holes. An X-ray outburst of KV UMa was detected from the RXTE satellite on 29 March 2000 \citep{remillard2000}. At the same time the optical brightness of the system had grown by $\approx 6^m$ from $V\approx 18.8^m$ in quiescence to $V\approx 12.9^m$ in the maximum \citep{uemura2000a,garcia2000}.

The galactic latitude of KV UMa is considerably high: $b=+62^{\circ}$. Along with the distance to the system $1.9\pm 0.4$ kpc \citep{wagner2001} it corresponds to a significant height over the galactic plane $z=1.7\pm 0.4$ kpc. The interstellar absorption to KV UMa is very weak, $E(B-V) = 0.013\div 0.017$, $A_v = 0.05^m$, it eases the multi-wavelength analysis of this system \citep{shahbaz2005,khargharia2013}.

Multivawelength observations of KV UMa during outbursts were conducted by \citet{hynes2003,torres2004}. They observed superhumps that indicated the precession of the accretion disk \citep{cook2000,uemura2000b,patterson2000,dubus2001}.

By the end of August 2000 the KV UMa brightness returned to the value before the outburst ($V\approx 19^m$). Spectroscopic observations during that period of time by \citet{wagner2001,mcclintock2001a} allowed: (i) to find spectral type the optical star (K7-M0)\text{V}, (ii) to obtain a reliable radial velocity curve, (iii) to calculate the orbital period, and (iv) to compute spectroscopic elements of the system. The KV UMa mass function turned out to be very high, $f_v(M) = 6.1\pm 0.3M_{\odot}$. The contribution of the optical star radiation in the total flux in the wavelength 5900\r{A} estimated using spectrophotometry in average was 32\%$\pm$6\%.

\citet{torres2004} showed that the evolution of the KV UMa spectrum in the wavelength range $\lambda=5800\div 6400$\r{A} from the beginning of the outburst decay until the quiescence took place due to the change of the contribution of the donor star in the total flux from 35\%$\pm$8\% on 09 December 2000 and on 26 January 2001 to 60\%$\pm$10\% on 02-03 January 2003. It is consistent with data from other authors: 53\%$\pm$7\% in April 2001 \citep{zurita2002} and 45\%$\pm$10\% in January 2002 \citep{mcclintock2003}. \citet{khargharia2013} gave a spectrophotometric estimate of the contribution of the optical star in the \textit{H} band as 54\%$\pm$27\% in April 2011, i.e. in quiescence.

Doppler tomography of KV UMa was made by \citet{torres2004,calvelo2009,zurita2016}. \citet{zurita2016} 
analysed movements of the  $H_{\alpha}$ emission centroid and found that the precession period of the accretion disk was $\approx 52$ days. It was in agreement with the precession period of the disk found using the analysis of superhumps in light curves \citep{cook2000,patterson2000,uemura2002,zurita2002}. Also a narrow $H_{\alpha}$ emission component was observed. It belonged to the optical star and, according to \citet{zurita2016}, it indicated a chromospheric activity of the donor star.

IR light curves in \textit{J}, \textit{K} bands were obtained in April 2003 and in March 2004 by \citep{mikolajewska2005}. In the \textit{J} band the star had the same brightness in minima, but maxima were not equal. The non-stellar component contribution in \textit{J}, \textit{K} bands in April 2003 and March 2004 did not exceed 33\% and 25\% respectively.

The first modelling of the KV UMa optical light curve was made by \citet{mcclintock2001b} under the assumption that the contribution of the accretion disk in the total brightness was 66\%, the inclination of the system's orbit was computed ($i=80^{\circ}$). The KV UMa light curve in the \textit{R} band was observed by \citet{wagner2001}. In the model with an accretion disk around the relativistic object (its contribution was 76\%) the inclination of the orbit was $i=81^{\circ}\pm 2^{\circ}$. That light curve was interpreted by \citet{khruzina2005} in the model that included the disk and the region of the interaction between the accretion stream and the disk (``hot line''). The inclination of the orbit was estimated to be $i=80^{+1}_{-4}$ degrees.

\citet{gelino2006} obtained \textit{BVRJHK} light curves of KV UMa in January 2003. Under the assumption about a negligible contribution of the accretion disk in IR wavelength (less than 8\%; light curves in IR are symmetric with practically equal maxima and minima) they computed the inclination of the orbit $i=68^{+2.8}_{-2}$ degrees and gave the estimate of the black hole mass $M_{BH}= 8.53M_{\odot}\pm 0.6M_{\odot}$. \citet{khargharia2013} conducted spectroscopic observations of KV UMa in the range $0.9\div 2.45$ micrometers and obtained a light curve in the \textit{H} band. As in the work by \citet{mikolajewska2005} their  \textit{H} light curve had different maxima and equal minima. \citet{khargharia2013} considered two models (a model with an accretion disk and a model with a disk and a hot spot on the outer border of the disk) and estimated the orbit's inclination as $68^{\circ}\leq i\leq 79^{\circ}$ and the black hole mass $6.9M_{\odot}\leq M_{BH}\leq 8.2M_{\odot}$. They emphasised that KV UMa  (similar to other X-ray novae in quiescence) demonstrated the presence of a continued activity even if the system was in its quiescent state with a very low X-ray luminosity.

We made long-lasting optical, \textit{J}, and \textit{K} photometric observations of KV UMa in order to find manifestations of such activity and to determine the inclination of the orbit and the black hole mass in the frames of an adequate model of the system. In addition it was planned to reconstruct the non-stellar component spectrum and to attempt to find traces of the precession of the elliptical accretion disk. Also some evolutionary calculations for low mass X-ray binaries with black holes were made.

Section 2 describes our observations of KV UMa conducted in 2017-2018 in infrared and optical ranges, telescopes and techniques used for them. Section 3 depicts light curves obtained from our observations and compares our curves with previous results of other authors. Section 4 presents a theoretical modelling of obtained light curves, Section 5 evaluates the model. Section 6 shows the results of the modelling, describes our spectrophotometric conclusions concerning the non-stellar component. Section 7 discusses the model and its applications. Appendix studies the evolution of low mass X-ray binaries with black holes paying special attention to the rapid decrease of the orbital period in closest pairs.

\section{Observations}

\begin{figure}
\center
\includegraphics[width=\columnwidth]{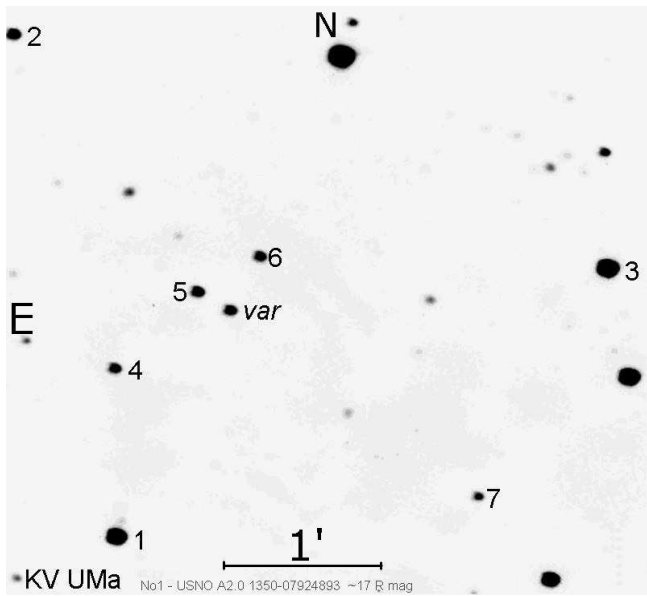}
\vspace{5pt} \caption{A finding chart for KV UMa surroundings, 4$'\times$4$'$, ``var'' is KV~UMa, the north is ``N'', the east is ``E''. }\label{figure1}
\end{figure}

\begin{table}
\large
\centering
\caption{A log of KV UMa observations.}
\label{obs-log}
\begin{tabular}{@{}cccc@{}}
\hline
Data & JD-2450000 & Band & Telescope \\
\hline
24 November 17 & 8082.622--.664 & \textit{C} & EMT \\
25 November 17 & 8083.458--.665 & \textit{C} & EMT \\
15 December 17 & 8103.411--.416 & \textit{J}, 100 s & CMO \\
16 January 18 & 8135.475--.541 & \textit{J}, 100 s & CMO \\
28 January 18 & 8147.426--.492 & \textit{K}, 30 s & CMO \\
28 January 18 & 8147.431--.494 & \textit{J}, 30 s & CMO \\
07 February 18 & 8157.378--.632 & \textit{J}, 30 s & CMO \\
07 February 18 & 8157.384--.631 & \textit{K}, 30 s & CMO \\
08 April 18 & 8217.313--.552 & \textit{J}, 100 s & CMO \\
17 April 18 & 8226.304--.400 & \textit{K}, 60 s & CMO \\
24 April 18 & 8233.300--.755 & \textit{K}, 60 s & CMO \\
25 April 18 & 8234.409--.560 & \textit{J}, 100 s & CMO \\
26 April 18 & 8235.374--.554 & \textit{J}, 100 s & CMO \\
01 June 18 & 8271.281--.380 & \textit{C} & EMT \\
04 June 18 & 8274.279--.378 & \textit{C} & EMT \\
05 June 18 & 8275.281--.785 & \textit{C} & EMT \\
06 June 18 & 8276.295--.372 & \textit{C} & EMT \\
07 June 18 & 8277.283--.371 & \textit{C} & EMT \\
09 June 18 & 8279.280--.365 & \textit{C} & EMT \\
01 November 18 & 8424.510--.566 & \textit{C} & EMT \\
02 November 18 & 8425.489--.637 & \textit{C} & EMT \\
03 November 18 & 8426.491--.641 & \textit{C} & EMT \\
04 November 18 & 8427.537--.642 & \textit{C} & EMT \\
05 November 18 & 8428.639--.623 & \textit{C} & EMT \\
08 November 18 & 8431.490--.646 & \textit{C} & EMT \\
09 November 18 & 8432.502--.625 & \textit{C} & EMT \\
10 November 18 & 8433.473--.639 & \textit{C} & EMT \\
13 November 18 & 8436.486--.534 & \textit{C} & EMT \\
\end{tabular}
\end{table}

Observations of KV UMa were performed during three seasons in 2017 and 2018 in optic and infrared spectral ranges. Since the orbital period of the system is short (about 4 hours), we tried to cover it completely during the night depending on weather conditions. At that time KV UMa was in quiescence. The observation log is given in Table \ref{obs-log}.

For the integral light (marked as ``C'') the 180 s exposition was used. In IR range it was from 30 to 100 seconds (see Table \ref{obs-log}) dependently on weather conditions.

\subsection{IR observations}

IR observations in \textit{J} ($\lambda_{eff}\approx 1.25$ $\mu$m) and \textit{K} ($\lambda_{eff}\approx 2.2$ $\mu$m) bands of the Mauna Kea Observatories (MKO) photometric system were conducted on newly installed 2.5-m telescope at Caucasian mountain observatory of SAI MSU (Sternberg Astronomical Institute, Lomonosov Moscow State University) located in Karachay-Cherkess Republic (Russian Federation) at the altitude 2112 meters above the sea level (CMO in Table \ref{obs-log}).

Infrared observations in \textit{J} band were conducted on 15 December 2017, 16 and 28 January, 7 February, 8, 25 and 26 April 2018, in \textit{K} band on 28 January, 7 February and 17 April 2018, a short piece of photometric observations was obtained in \textit{H} band on 18 January 2018. The ASTRONICAM camera-spectrograph \citep{nadjip2017} with the Hawaii-2RG detector (2048$\times$2048 pixels) was used, in the photometric regime only the central part (1024$\times$1024 pixels) of it can be in operation. 

2MASS J11181198+4802190 was used as a comparison star for IR observations (``5'' in Fig. \ref{figure1}). We attributed following MKO magnitudes for it: $J=16.247$, $H=15.624$, $K=15.46$.

An average error in IR bands estimated using control stars 4 and 6 (see Fig. \ref{figure1}) was $0.02^m$. During the first analysis of data in \textit{J} band we divided them in two series: from 15 December 2017 to 07 February 2018, and from 08 to 26 April 2018. The dimension of the first series was 298 points, for the second it was 404 points, as the result of averaging of individual measurements we got 60 and 81 points respectively. It was found that light curves changed insignificantly during observations, so we used combined series to create an averaged \textit{J} light curve.
     
In the \textit{K} band there were obtained 415 points, then the mentioned 5 image-averaging were performed to increase the precision. After removing points out of $3\sigma$ level (as for the \textit{J} band), 83 points remained (28 January 2017 --- 25 April 2018). The error in the \textit{K} band is higher than in the \textit{J} band and equals to $0.03^m$ (in average). 

\subsection{Optical observations}

Optical observations were unfiltered and performed in the integral light with the effective wavelength $\lambda_{eff}\approx 6400$\r{A} that corresponded to the average wavelength with the bandwidth at half intensity $\lambda=4300\div 8300$\r{A} (see, e.g., \citealp{armstrong2013,khruzina2015}, where parameters of this band were discussed) using 1.25 m V. P. Engelgardt Mirror Telescope at the Crimea Astronomical Station of M. V. Lomonosov Moscow State Univeristy (EMT in Table \ref{obs-log}) with the CCD camera VersArray-1300. The object was observed on 24 and 25 November 2017, 1, 2, 4-7, 9  June, 1-5, 8-10, 13 November 2018.
 
To process optical data the USNO A2.0 1350-0792893 reference star was used with following stellar magnitudes: $B=19.144$, $V=17.758$, $R_c=16.895$, $I_c=16.352$ (according to observational data in November 2018), where $R_c$ and $I_c$  were in the Cousins system. Also we conducted quasi-synchronous observations in integral light (\textit{C}) and in $R_c$ band. A comparison of stellar magnitudes $C=18.821$ and $R_c=18.820$ showed that it was possible to use $R_c$ value to make a light curve. 
     
Bias and flat field corrections were made for all \textit{C}, \textit{J}, and \textit{K} images.

\section{Light curves}

\begin{figure}
\renewcommand{\thefigure}{\arabic{figure}a}
\center
\includegraphics[width = \columnwidth]{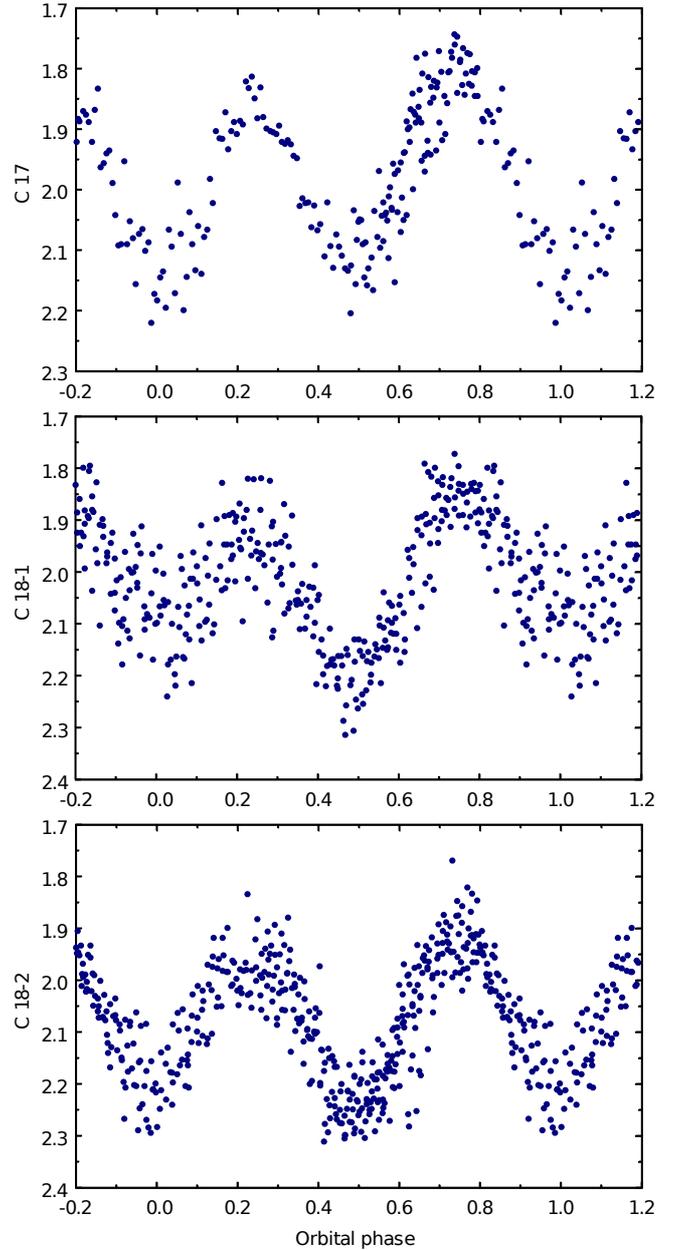}
\vspace{5pt} \caption{Average phase optical light curves of KV UMa obtained during the season (see Table \ref{obs-log}). The data were convolved using Ephemeris \ref{eph}.}
\label{figure2a}
\end{figure}

\begin{figure}
\addtocounter{figure}{-1}
\renewcommand{\thefigure}{\arabic{figure}b}
\center
\includegraphics[width = \columnwidth]{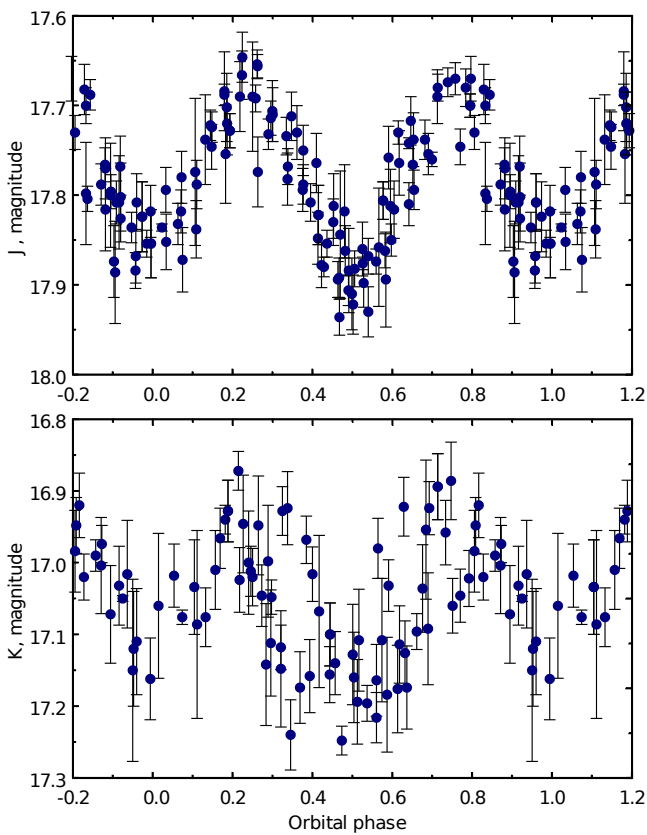}
\vspace{5pt} \caption{Average IR phase light curves of KV UMa obtained during the season (see Table \ref{obs-log}). The data were convolved using Ephemeris \ref{eph}.}
\label{figure2b}
\end{figure}

To compute light curves we used following ephemeris:

\begin{equation}
\label{eph}
\textrm{Min}\ (\varphi=0.0) = T_0+P_{orb}\times E.                                      
\end{equation}

\noindent where $T_0=\textrm{HJD}\ 245\ 5676.6017$ is the initial epoch, $P_{orb}=0.1699337\pm 0.0000002$ days, $E$ is the number of orbital cycles after $T_0$.

The $\varphi=0.0$ phase corresponds to the upper conjunction of the relativistic object (the optical star is in front of it; \citealp{g-h-2014}).

Fig. \ref{figure2a} shows photometric measurements in white colour convolved with the orbital period ($\Delta C$ is the difference of stellar magnitudes with respect to the main comparison star as $\Delta C = C_{var}-C_{control}$), and Fig. \ref{figure2b} is the same for \textit{J}, \textit{K} bands. Our measurements are shown for three seasons in white colour in November 2017 (JD2458082--8083), June 2018 (JD2458271--8279), November 2018 (JD2458424--8436), in \textit{J}, \textit{K} from the end of December 2017 to April 2018 (JD2458103--8235).

\begin{figure}
\center
\includegraphics[width=\columnwidth]{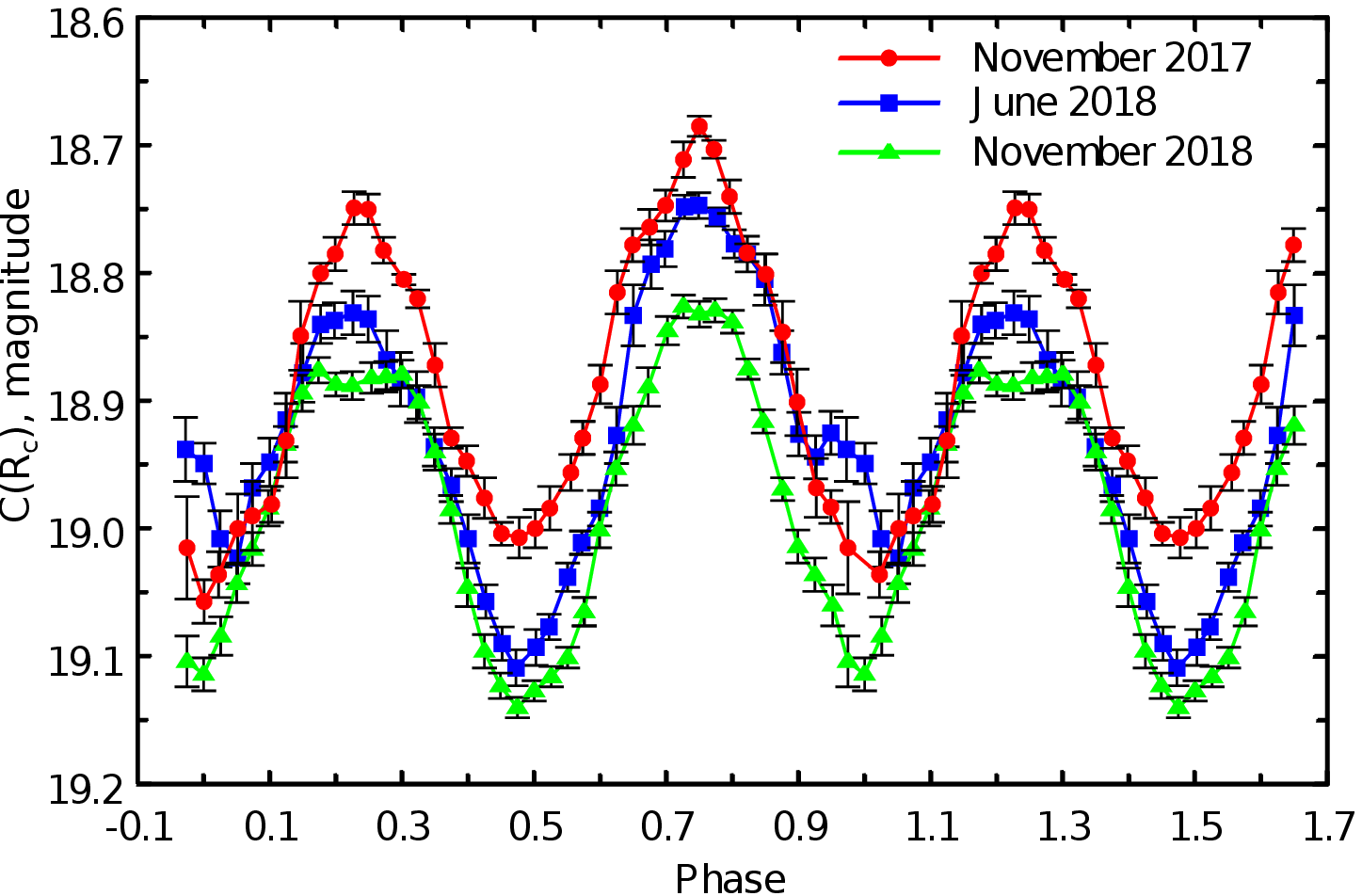}
\vspace{5pt}
\caption{Average phase curves of KV UMa in the integral light. About $\sim 1000$ individual measurements were used. Errors of individual observations were $0.025^m\div 0.030^m$ (in average).}
\label{figure3}
\end{figure}

We did not detect the presence of a considerable flickering within light curves during three observational seasons (about 350 days dependently on the wavelength band), see Table \ref{obs-log}. The scatter of individual points can be basically explained by errors of measurements estimated using control stars. The average optical brightness of the system changed significantly (see Fig. \ref{figure3}, light curves obtained in November  2017, June 2018, and November 2018), the system monotonically faded. It is clearly seen in the composite image (Fig. \ref{figure3}) that shows three average phase curves in the integral light in November 2017 ($C17$), June and November 2018 ($C$18-1 and $C$18-2). The number of images obtained in November 2017 was $N=220$, an average stellar magnitude of the system $C=18.86\pm 0.01$, in June 2018 $N=327$, $C=18.91\pm 0.01$, in November 2018 $N=434$, $C=18.97\pm 0.01$.  The averaged integral stellar magnitude in November 2017 -- June 2018 was $C=18.929\pm 0.004$. 

Fig. \ref{figure3} shows that light curves in integral light changed during the year, secondary maxima became lower by about $0.15^m$, the shape of the primary minimum in November 2018 also changed (it became flat). Optical light curves showed considerable inequality of maxima with the same character (but expressed less) as IR light curves obtained by \citet{mikolajewska2005,khargharia2013}. Optical light curves also showed inequality in minima that changed its sign in different epochs: in November 2017 the system was brighter in the secondary minimum than in the primary minimum ($\varphi=0.0$). In June and November 2018 the situation was opposite: the system was brighter in the primary minimum than in the secondary minimum (see Fig. \ref{figure3}). In IR (Fig. \ref{figure2b}) the inequality of maxima had the opposite sign in comparison with the optical range: the maximum at the phase $\varphi=0.25$ was higher that at the phase $\varphi=0.75$.

A comparison of our IR data with data by \citet{mikolajewska2005} obtained in April 2003 and March 2004 showed that our light curve in \textit{J} band had different ratio of maxima (the maximum at the phase 0.25 was higher than at the phase 0.75). In addition the average brightness and the colour index of the system were $J=18.03^m\pm 0.04^m$, $J-K=1.11^m\pm 0.2^m$ in the paper by \citet{mikolajewska2005}, in our case they were $J= 17.79^m\pm 0.02^m$, $J-K=0.73^m\pm 0.04^m$, i.e. the system became brighter in \textit{J} band by $0.23^m$ and bluer. The amplitude of the orbital variability in the \textit{J} band in our case dropped to $\Delta J\approx 0.23^m$ in comparison to a \textit{J} light curve by \citet{mikolajewska2005} ($\Delta J\approx 0.35^m$). Those differences were most likely connected with the variability of the contribution of the non-stellar component to the total brightness of the system (the accretion disk with the region of the interaction between the disk and the gas stream). Despite of different photometric system the difference in the brightness in 2003-2004 and 2017-2018 should be connected with the physical variability.
     
It should be emphasised that in the optical light and in the IR range a considerable growing of the average brightness of the system was not accompanied by the increase of the flickering. In this feature KV UMa differed from the low mass X-ray nova A0620-00 (see, e.g., \citealp{shugarov2016,cherepashchuk2019}).

\section{Modelling of light curves}

\begin{figure*}
\renewcommand{\thefigure}{\arabic{figure}a}
\center
\includegraphics[width = \textwidth]{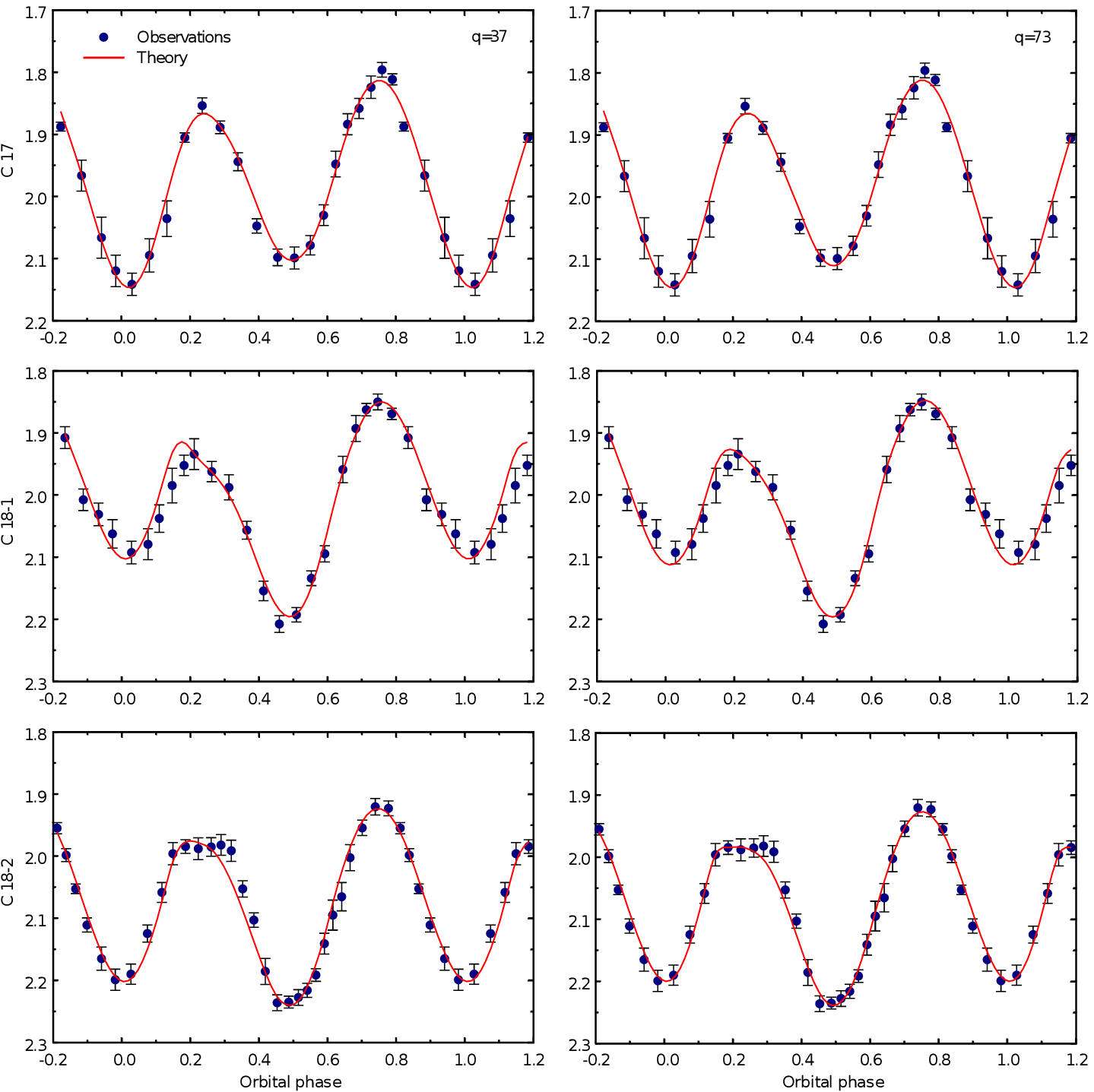}
\vspace{5pt} \caption{Average light curves of KV UMa and optimal theoretical light curves in the integral light for parameters from Tables \ref{table2a} and \ref{table2b}; two values of the mass ratio $q=M_{BH}/M_v =37$ and 73 were applied.}
\label{figure4a}
\end{figure*}

\begin{figure*}
\addtocounter{figure}{-1}
\renewcommand{\thefigure}{\arabic{figure}b}
\center
\includegraphics[width = \textwidth]{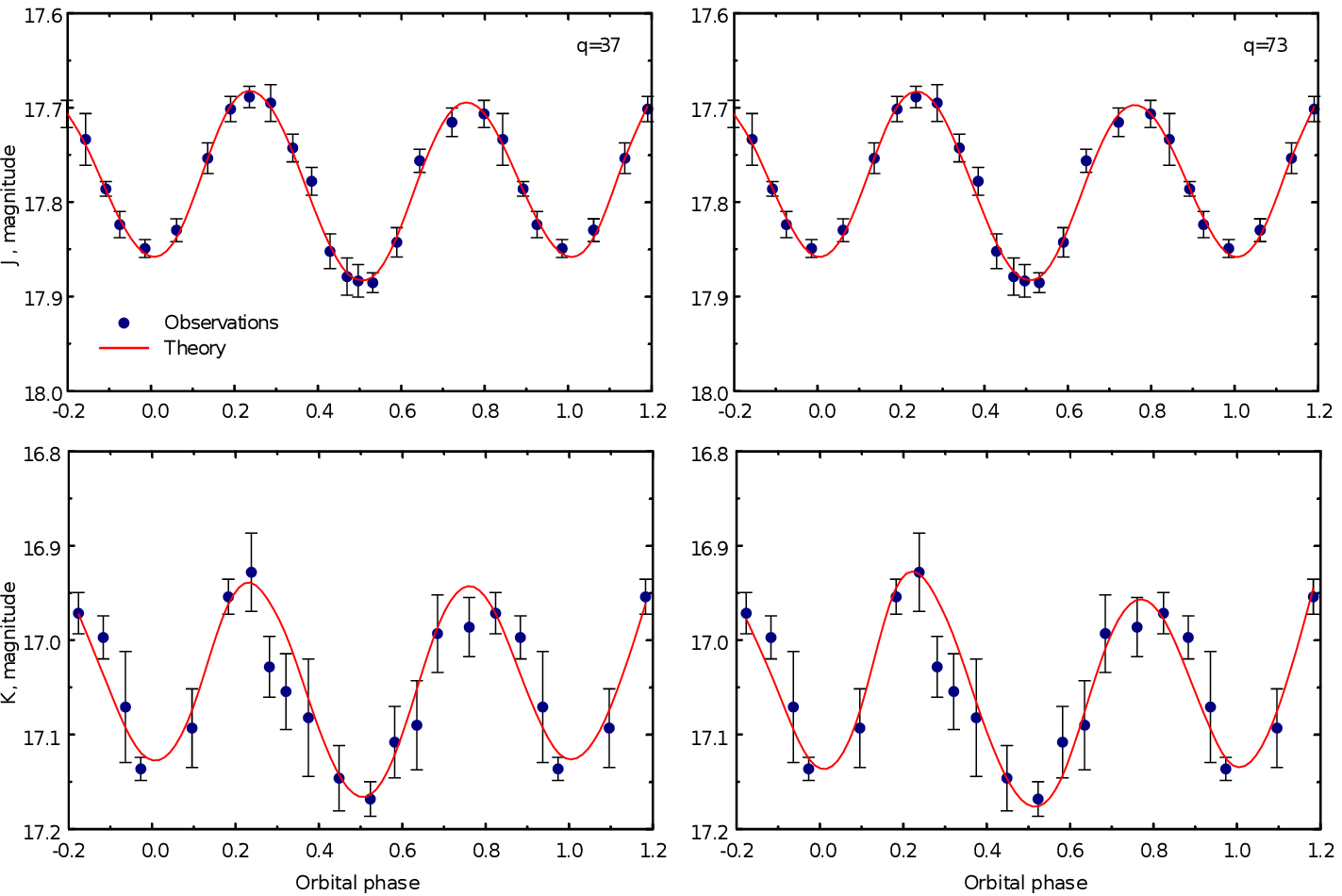}
\vspace{5pt} \caption{Average light curves of KV UMa and optimal theoretical light curves in IR for parameters from Tables \ref{table2a} and \ref{table2b}; two values of the mass ratio $q=M_{BH}/M_v =37$ and 73 were applied.}
\label{figure4b}
\end{figure*}

Average standard light curves were used for the modelling. The orbital period was divided by phase intervals, in each interval a mean of individual stellar magnitudes was computed along with its mean square error.

Figs. \ref{figure4a} and \ref{figure4b} shows average optical and IR light curves (dots) and optimal theoretical light curves (lines) that correspond to final optimal values of parameters (see Tables \ref{table2a} and \ref{table2b}) for two values of the mass ratio $q=M_{BH}/M_v$, where $M_{BH}$ and $M_v$ are masses of the black hole and the optical star respectively. 

For the modelling we used a model of an interacting binary system. The model had already been successfully applied to analyse light curves of cataclysmic binary systems (see, e.g., \citealp{khruzina2001,khruzina2003a,khruzina2003b}, a detailed description was done by \citealp{khruzina2011}) along with X-ray binary systems \citep{khruzina2005,cherepashchuk2019}. In this case the standard method to synthesise light curves was used \citep{wilson1971}. The optical star fills its Roche lobe. A gravitational darkening and a limb darkening were taken into account (a liner limb darkening law). An elliptical accretion disk was in the system, a relativistic companion was in one of its focuses. Near the outer border of the disk there was a region of an interaction of a gas stream  (a hot line and a hot spot). The hot line was located along of the gas stream, the hot spot was located on the outer border of the disk. A heating of the hot line arises due to the lateral collision of the matter of the stream with the rotating matter of the halo around the disk. The matter heated in corresponding shock waves cooled, joined the outer border of the disk and formed the hot spot. This feature differs our model from the classical model with a hot spot in which the spot on the outer border of the disk is heated due to the frontal collision of the stream and the disk. Our model with the hot line and the hot spot satisfies the results of three dimensional hydrodynamical calculations by \citet{bisikalo2005,lukin2017}, who showed that the interaction of the stream with the rotating disk in the frames of 3D model occurs in a more complicated way than in the classical model of the hot spot.

It is necessary to note that it is very difficult to distinguish the classical model with the hot spot and our model with the interaction region using Doppler tomography, because the widely used method of the maximum entropy to solve ill posed problems allows to find a stable approximate solution with a minimal tiny structure. The method of the maximum entropy (see, e.g., \citealp{tikhonov1983,marsh1988}) gives smoothed results.

In general form our model can be described using twenty parameters, they can be found if eclipses exist in the cataclysmic binary. In KV UMa eclipses do not observed, therefore we restricted a group of parameters with limited values and fixed values of other parameters. The technique of calculations can be found in our previous work \citep{cherepashchuk2019}.

Two values of components mass ratio were used: $q=M_v/M_{BH}=37$  \citep{g-h-2012} and $q=73$ \citep{petrov2017}, they were estimated using the rotational broadening of line profiles in the donor star spectrum. This allowed to test the sensitivity of the problem to the change of $q$. The temperature of the optical K7V star was fixed as $T_2=4120$ K. Fluxes from elementary areas on the star, on the disk and within the interaction region were computed using Planck's law. Ranges of permitted values of other parameters were close to ranges used for V616 Mon by \citet{cherepashchuk2019}.
As the result in our calculation remained 11 free parameters: $i$, $R_d/ \xi$ ($\xi$ is the the distance between the inner Lagrange point L1 and the black hole), the disk eccentricity $e$, the azimuth of the periastron of the disk $\alpha_e$, parameters $T_{in}$ and $\alpha_g$ that characterize the distribution of the temperature in the disk

$$
T(r)=T_{in}\left(\frac{R_1}{r}\right)^{\alpha_g},
$$

\noindent where $R_1=0.0003a_0$ ($a_0$ is the radius of the relative orbit), $a_v/a_0$, $b_v/a_0$ are the semi-major and semi-minor axes of the ellipsoid that fits the hot line in units of $a_0$, $T^{(1)}_{max}$, $T^{(2)}_{max}$ are the maximum temperatures in the ``front'' (windward) side and in the ``far'' (leeward) side of hot line with respect to the direction of the disk rotation, $R_{sp}/a_0$ is the semi-major axis of the hot spot ellipse in units of $a_0$. Values of $e$ and $\alpha_e$ were not fixed to let to see the dynamics of the changes of these parameters with time. The solution of the inverse problem was conducted using iteration over the parameter $i$.  

For every fixed value of $i$ the minimization of the residual functional over all other ten parameters was realized. We call here ``the residual functional'' the weighted sum of squares of differences between the observed light curve an the theoretical one. The Nelder-Mead method was used to find the minimum of residuals \citep{himmelblau1972}. 

\section{An adequacy test for the model}

To test the adequacy of the model (including estimations concerning contributions of different radiating structures) let us consider our average optical light curve obtained in November 2017. The adequacy of our model for other epochs was studied too and it was illustrated by Figs. \ref{figure8a} and \ref{figure8b}.

\begin{figure}
\center
\includegraphics[width = \columnwidth]{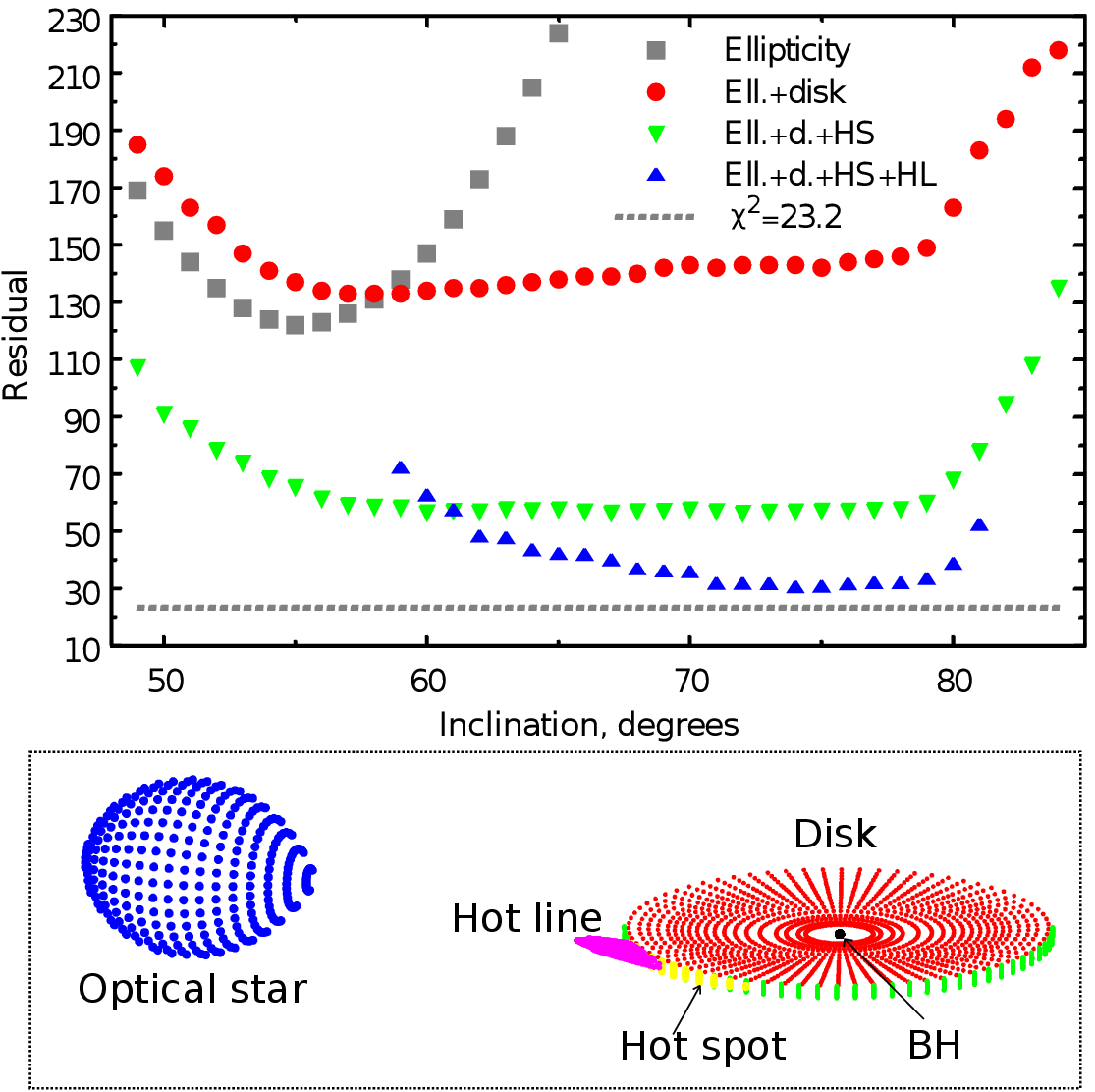}
\vspace{5pt} \caption{The adequacy test for models for the $C17$ light curve. The upper panel presents dependencies of residual on the inclination of the orbit $i$ that are minimal over all other parameters, $q=37$: ``ell'' is a ``purely'' ellipsoidal model, there is no disk with an interaction region; ``ell+disk'' is the model with the ellipsoidal donor star and the disk without a hot spot and a hot line; ``ell+disk+HS'' is the model with the ellipticity effect and with a disk with a hot spot; ``ell+disk+HS+HL'' is the model that includes the effect of ellipsoidal star, the accretion disk around the black hole with the hot spot and the hot line. The horizontal line corresponds to the critical value $\chi^2_{n-m,0.01}$ within 1\% confidence level. The bottom panel presents a schematic picture of the used model.}
\label{figure5}
\end{figure}

Fig. \ref{figure5} shows curves of residuals as functions of the orbital inclination $i$ with minimal values of residuals over all other parameters in models of the system. For each $i$ value there is a definite corresponding average value of the non-stellar component luminosity (the disk plus the hot spot plus the hot line), which was determined by the solution of the inverse problem. In Fig. \ref{figure5} there is a model of the system ($\varphi=0.695$) that includes the donor star, the disk with the hot spot, and the hot line for optimal values of parameters ($C17$ curve) and $i=74^{\circ}$. The dashed line cuts the critical value of $\chi^2=23.2$ within the significance level 1\% for the degree of freedom $n-m=12$ (in the model ``ell+disk+HS+HL'', $n=22$ is the number of points in the light curve, $m=10$ is the number of parameters for the minimization).

\begin{figure}
\center
\includegraphics[width = \columnwidth]{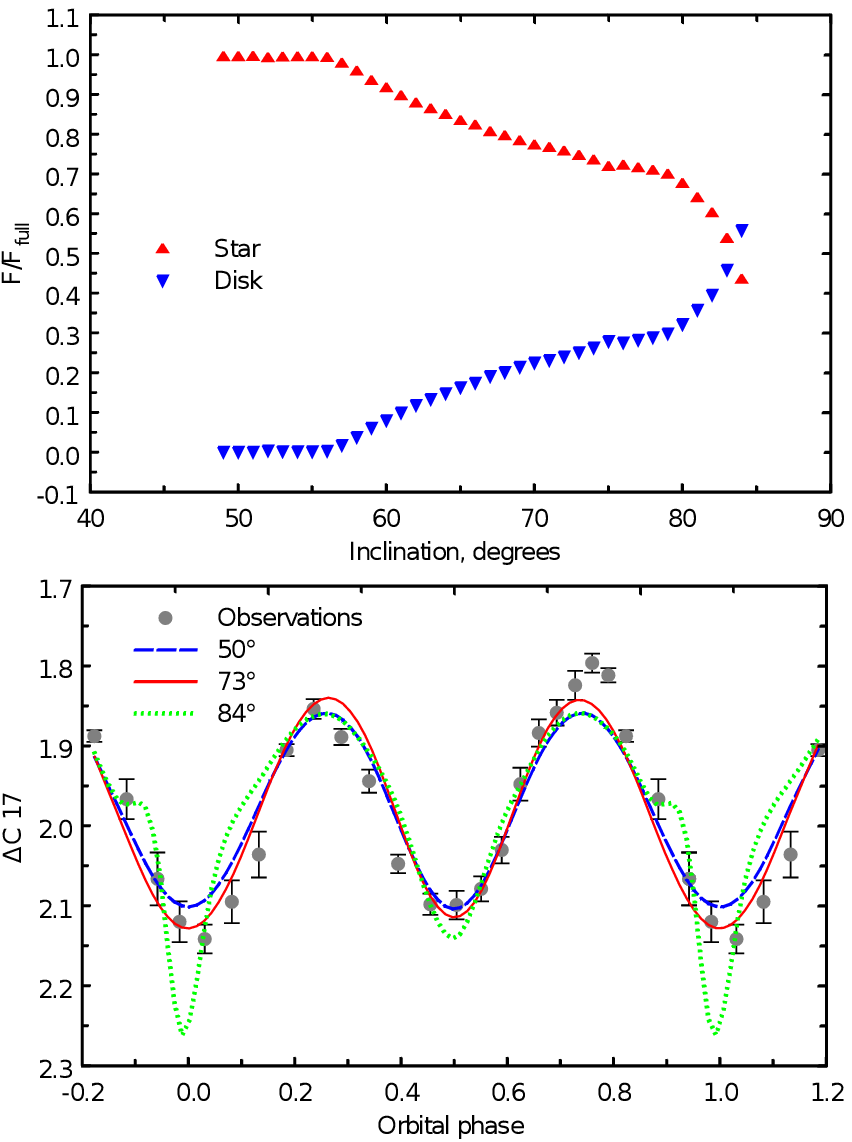}
\vspace{5pt} \caption{ An output of the model ``an ellipticity effect plus a disk without a hot spot and a hot line'' for the average optical light curve $\Delta C17$ (November 2017) and $q=37$ (see Fig. \ref{figure5}). The upper panel shows the contribution of the star and the disk to the total system's luminosity depending on the inclination $i$. The lower panel shows the average light curve (dots) and optimal theoretical curves for $i=50^{\circ}$, $73^{\circ}$, $84^{\circ}$. For $i=50^{\circ}$ the disk contribution is zero. For $i=84^{\circ}$ the contribution is maximal ($\approx 55$\%), the disk is eclipsed by the donor star. }
\label{figure6}
\end{figure}

\begin{figure}
\center
\includegraphics[width = \columnwidth]{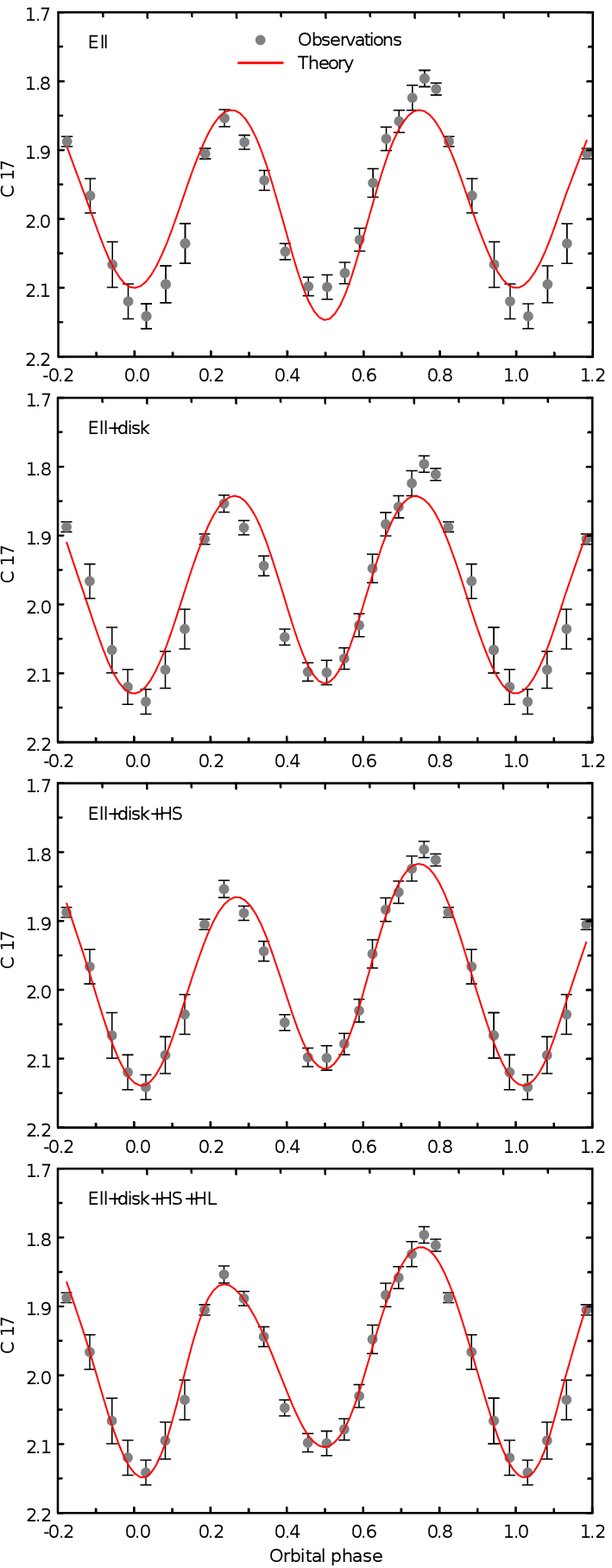}
\vspace{5pt} \caption{An illustration for resuduals from Fig. \ref{figure5}. The average optical light curve of KV UMa (November 2017) is shown along with optimal theoretical light curves that correspond to four models of the system: ``ell'' ($i=55^{\circ}$), ``ell+disk'' ($i=57^{\circ}$), ``ell+disk+HS'' ($i=67^{\circ}$), ``ell+disk+HS+HL'' ($i=74^{\circ}$).}
\label{figure7}
\end{figure}

It can be seen that for the tidally deformed optical star only (see for the ellipticity effect the paper by \citealp{lyutyi1973}) without the disk, the hot spot, and the hot line the corresponding curve of residuals has a clear minimum around $i\approx 55^{\circ}$. However, the minimal value of the residual  ($\chi^2_{min}= 122$) is more than three times higher than the critical value $\chi^2_{n-m,0.01}=38.9$ for this model ($m=1$). So, the ``pure'' ellipsoidal model is surely rejected. If we take into account the accretion disk (without the hot spot and the hot line) along with the donor star the value of the residual changes purely. This means that this model is also rejected, the minimum in the residual curve becomes wide and flat, because the ellipticity effect can be compensated with the change in the disk luminosity (see Fig. \ref{figure6}). The spectrophotometric estimate of the non-stellar component allows to choose a point inside the flat minimum corresponding to the optimal value of $i$. The fact that both described models (the only ellipsoidal star and the ellipsoidal star plus the accretion disk around the black hole) are strongly inadequate to observations is connected with a significant inequality of maxima in quadratures of the light curve (see Fig. \ref{figure7}) that can not be described in these models.

In the model with the donor star, the disk, and the hot spot on the outer border of the disk the minimal value of the residual drops twice ($\chi^2_{min}=56.4$, $\chi^2_{n-m,0.01}=34.8$, $m=4$), but it is still significantly higher than the critical value $\chi^2$, i.e. this model is also surely rejected.

The residual curve in this case also has a wide and flat minimum ($i=58\div 79^{\circ}$), and to find $i$ it is necessary to independently know the non-stellar component luminosity.

In the model with the donor star, and with the disk with the hot spot and the hot line the minimal value of the residual ($\chi^2_{min}=30.2$, $\chi^2_{n-m,0.01}=26.2$, $m=10$) is $\approx$15\% higher than the critical value $\chi^2\approx 26.2$ within the confidence level $\alpha=0.01$ and it is lower than the critical value for $\alpha=0.001$ ($\chi^2_{12,0.001}= 32.9$). So, in this case there is a more or less good basis to accept the model. The minimum of residuals determines the optimal value of $i=74^{\circ}$, the corresponding average luminosity of the non-stellar component equals to 0.38 of the total flux from the system, the last theoretical value can be controlled with a spectrophotometric estimate of this luminosity. In addition luminosities of both components (the disk plus the hot spot and the hot line) can be found. As follows from Fig. 7 in this case it is possible to make a good description for the amplitude of the light curve and for the inequality of brightness maxima  in quadratures.

So, with some reservations, it is possible to accept that the model ``the star plus the accretion disk with the hot spot and the hot line'' is adequate to observational data, and therefore this model will be used to find the orbital inclination of KV UMa and the black hole mass using the wavelength range $\lambda=6400\div 22000$\r{A}.

\section{Results of modelling}

\begin{figure*}
\renewcommand{\thefigure}{\arabic{figure}a}
\center
\includegraphics[width = \textwidth]{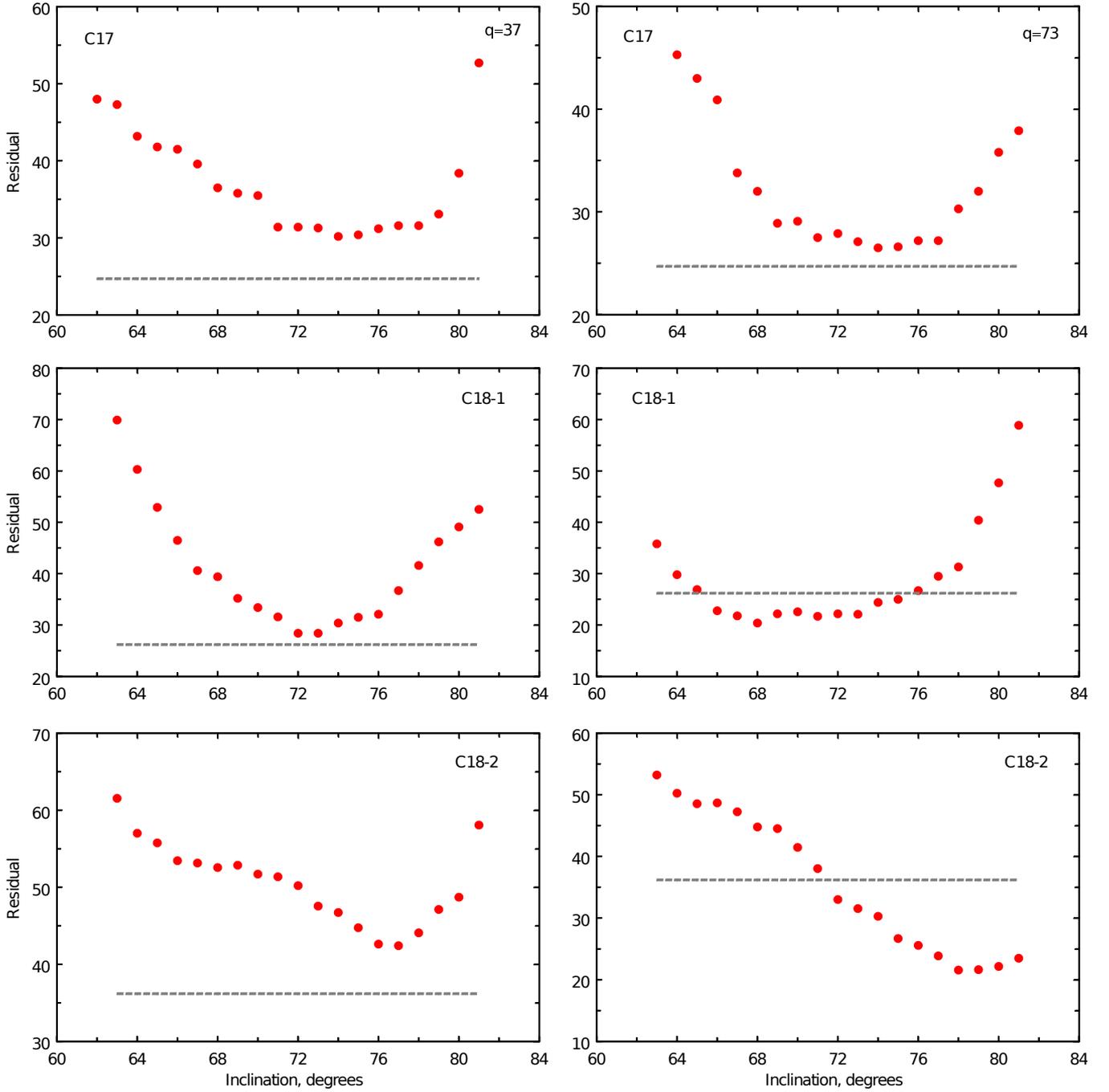}
\vspace{5pt} \caption{Curves of residuals between observational and theoretical optical light curves minimized over appropriate parameters as functions of the orbit's inclination $i$: a horizontal dashed line ``cuts'' the critical value of $\chi^2$ within the 1\% confidence level (see Tables \ref{table2a} and \ref{table2b}). Curves are shown for two component's mass ratios: $q=M_{BH}/m_v=37$ and 73.}
\label{figure8a}
\end{figure*}

\begin{figure*}
\addtocounter{figure}{-1}
\renewcommand{\thefigure}{\arabic{figure}b}
\center
\includegraphics[width = \textwidth]{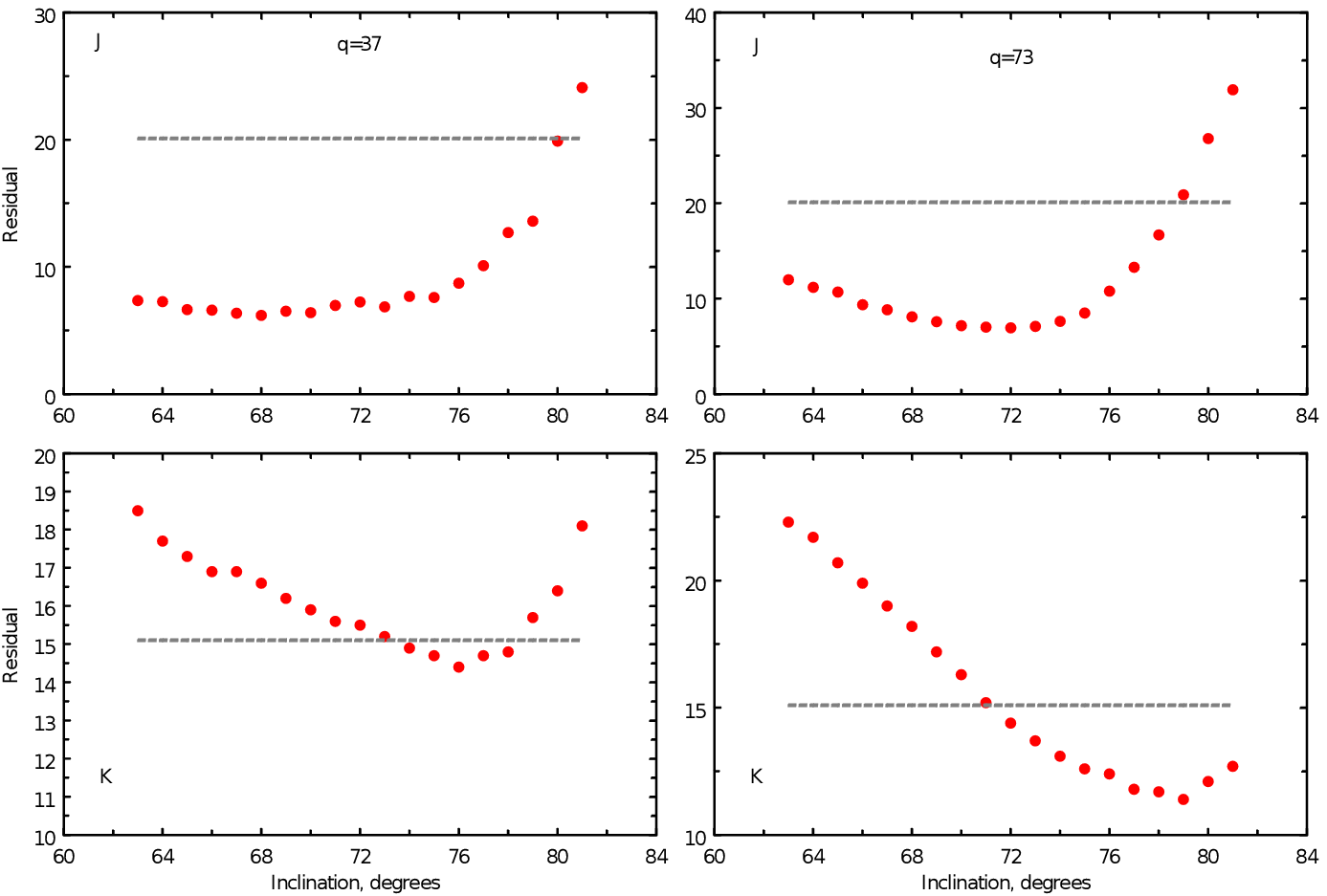}
\vspace{5pt} \caption{The same as Fig. \ref{figure8a} for infrared light curves.}
\label{figure8b}
\end{figure*}

Figs. \ref{figure8a} and \ref{figure8b} show minimized residuals as functions of the orbital inclination $i$ for two ratios of masses of components $q=M_{BH}/M_v=37$ and 73 in different epochs. Horizontal dashed lines ``cut'' the critical $\chi^2$ within the significance level 1\%. It can be seen that in most cases minimal values of residuals are below the critical value, therefore we accept that the model ``the donor star plus the accretion disk with the hot spot and the hot line'' is adequate to observations and there is no point to reject it. Using minima of residuals one can find optimal values of $i$ and corresponding parameters of the model. These values are shown in Tables \ref{table2a} and \ref{table2b}. It is evident that for $q=37$ and 73 the optimal value of $i$ found independently using five light curves (3 optical and 2 infrared) is in ranges $68^{\circ}\div 77^{\circ}$ and $68^{\circ}\div 79^{\circ}$ respectively. So, the optimal value of $i$ weakly depends on the mass ratio $q$. Combining data for two values of $q$ we make the final estimate of the inclination of the KV UMa orbit $i=74^{\circ}\pm 4^{\circ}$. 

\begin{table*}
\renewcommand{\thetable}{\arabic{table}a}
\large
\centering
\caption{The solution of the inverse problem of interpretation of optical and IR light curves for $q=M_{BH}/M_2=37$.}
\label{table2a}
\begin{tabular}{@{}cccccc@{}}
\hline
Parameters & $C17$ & $C18-1$ & $C18-2$ & $J$ & $K$ \\
\hline
JD 2450000+ & 8082-8083 & 8271-8279 & 8424-8436 & 8103-8235 & 8147-8234 \\
$N$ & 196 & 327 & 434 & 121 & 83 \\
$i$, $^{\circ}$ & 74 & 73 & 77 & 68 & 76 \\
\multicolumn{6}{c}{The disk with the hot spot} \\ 
$R_d$, $\xi$ & $0.394\pm 0.002$ & $0.398\pm 0.002$ & $0.401\pm 0.002$ & $0.385\pm 0.007$ & $0.366\pm 0.009$ \\
$a$, $a_0$ & $0.310\pm 0.002$ & $0.315\pm 0.004$ & $0.317\pm 0.002$ & $0.304\pm 0.006$ & $0.289\pm 0.009$ \\
$e$ & $0.025\pm 0.004$ & $0.019\pm 0.004$ & $0.022\pm 0.003$ & $0.025\pm 0.009$ & $0.03\pm 0.01$ \\
$\alpha_e$, $^{\circ}$ & $125\pm 2$ & $95\pm 2$ & $145\pm 15$ & $125\pm 7$ & $125\pm 25$ \\
$T_{in}$, K & $139010\pm 575$ & $122760\pm 210$ & $125780\pm 370$ & $114470\pm 265$ & $106775\pm 635$ \\
$\alpha_g$ & $0.692\pm 0.002$ & $0.660\pm 0.001$ & $0.657\pm 0.001$ & $0.622\pm 0.001$ & $0.605\pm 0.001$ \\
$R_{sp}$, $a_0$ & $0.24\pm 0.03$ & $0.28\pm 0.03$ & $0.25\pm 0.05$ & $0.13\pm 0.03$ & $0.09\pm 0.02$ \\
$0.5\beta_d$, $^{\circ}$ & $1.66\pm 0.05$ & $2.0\pm 0.1$ & $1.6\pm 0.1$ & $2.2\pm 0.1$ & $2.7\pm 0.2$ \\
\multicolumn{6}{c}{The hot line} \\ 
$a_v$, $a_0$ & $0.035\pm 0.001$ & $0.032\pm 0.001$ & $0.032\pm 0.001$ & $0.029\pm 0.001$ & $0.027\pm 0.003$ \\
$b_v$, $a_0$ & $0.257\pm 0.002$ & $0.215\pm 0.002$ & $0.233\pm 0.001$ & $0.261\pm 0.003$ & $0.275\pm 0.009$ \\
$T_{ww,max}$, K & $12370\pm 305$ & $16740\pm 250$ & $14935\pm 500$ & $18715\pm 500$ & $22090\pm 995$ \\
$T_{lw,max}$, K & $10060\pm 55$ & $12560\pm 40$ & $11410\pm 70$ & $13260\pm 100$ & $14775\pm 450$ \\
$F_d/F_{full}$ & $0.246\pm 0.003$ & $0.312\pm 0.003$ & $0.277\pm 0.002$ & $0.435\pm 0.003$ & $0.339\pm 0.001$ \\
$F_{HL}/F_{full}$ & $0.136\pm 0.009$ & $0.09\pm 0.01$ & $0.073\pm 0.008$ & $0.078\pm 0.005$ & $0.061\pm 0.007$ \\
$(F_d+F_{HL})/F_{full}$ & $0.382\pm 0.010$ & $0.40\pm 0.01$ & $0.350\pm 0.008$ & $0.513\pm 0.005$ & $0.401\pm 0.007$ \\
$n$ & 22 & 23 & 30 & 19 & 16 \\
$\chi^2$; $\chi^2_{crit}$ & 30.2; 24.7 & 28.4; 26.2 & 42.4; 36.2 & 6.19; 20.1 & 14.4; 15.1 \\
\end{tabular}
\end{table*}

\begin{table*}
\addtocounter{table}{-1}
\renewcommand{\thetable}{\arabic{table}b}
\large
\centering
\caption{The solution of the inverse problem of interpretation of optical and IR light curves for $q=M_{BH}/M_2=73$.}
\label{table2b}
\begin{threeparttable}
\begin{tabular}{@{}cccccc@{}}
\hline
Parameters & $C17$ & $C18-1$ & $C18-2$ & $J$ & $K$ \\
\hline
JD 2450000+ & 8082-8083 & 8271-8279 & 8424-8436 & 8103-8235 & 8147-8234 \\
$N$ & 196 & 327 & 434 & 121 & 83 \\
$i$, $^{\circ}$ & 74 & 68 & 78 & 72 & 79 \\
\multicolumn{6}{c}{The disk with hot spot} \\ 
$R_d$, $\xi$ & $0.398\pm 0.002$ & $0.403\pm 0.002$ & $0.367\pm 0.002$ & $0.379\pm 0.002$ & $0.362\pm 0.006$ \\
$a$, $a_0$ & $0.328\pm 0.001$ & $0.333\pm 0.002$ & $0.303\pm 0.001$ & $0.312\pm 0.001$ & $0.298\pm 0.005$ \\
$e$ & $0.025\pm 0.001$ & $0.021\pm 0.004$ & $0.022\pm 0.004$ & $0.025\pm 0.002$ & $0.025\pm 0.007$ \\
$\alpha_e$, $^{\circ}$ & $125\pm 2$ & $90\pm 18$ & $146\pm 5$ & $125\pm 1$ & $125\pm 30$ \\
$T_{in}$, K & $140075\pm 595$ & $129050\pm 495$ & $104000\pm 220$ & $110295\pm 265$ & $120070\pm 955$ \\
$\alpha_g$ & $0.721\pm 0.001$ & $0.716\pm 0.001$ & $0.624\pm 0.001$ & $0.613\pm 0.001$ & $0.593\pm 0.002$ \\
$R_{sp}$, $a_0$ & $0.27\pm 0.03$ & $0.22\pm 0.03$ & $0.19\pm 0.05$ & $0.08\pm 0.01$ & $0.10\pm 0.04$ \\
$0.5\beta_d$, $^{\circ}$ & $1.25\pm 0.01$ & $1.7\pm 0.1$ & $1.6\pm 0.1$ & $1.8\pm 0.1$ & $2.8\pm 0.1$ \\
\multicolumn{6}{c}{The hot line} \\ 
$a_v$, $a_0$ & $0.038\pm 0.001$ & $0.037\pm 0.001$ & $0.016\pm 0.001$ & $0.0294\pm 0.0001$ & $0.028\pm 0.003$ \\
$b_v$, $a_0$ & $0.265\pm 0.001$ & $0.236\pm 0.001$ & $0.184\pm 0.001$ & $0.280\pm 0.002$ & $0.298\pm 0.003$ \\
$T_{ww,max}$, K & $12130\pm 220$ & $13055\pm 340$ & $14125\pm 935$ & $16365\pm 450$ & $22725\pm 615$ \\
$T_{lw,max}$, K & $5285\pm 35$ & $10410\pm 45$ & $10325\pm 95$ & $11890\pm 100$ & $15350\pm 305$ \\
$F_d/F_{full}$ & $0.215\pm 0.004$ & $0.243\pm 0.003$ & $0.379\pm 0.001$ & $0.497\pm 0.005$ & $0.475\pm 0.002$ \\
$F_{HL}/F_{full}$ & $0.179\pm 0.009$ & $0.152\pm 0.013$ & $0.027\pm 0.005$ & $0.083\pm 0.006$ & $0.085\pm 0.011$ \\
$(F_d+F_{HL})/F_{full}$ & $0.394\pm 0.011$ & $0.395\pm 0.013$ & $0.406\pm 0.005$ & $0.580\pm 0.006$ & $0.560\pm 0.012$ \\
$n$ & 22 & 23 & 30 & 19 & 16 \\
$\chi^2$; $\chi^2_{crit}$ & 26.5; 24.7 & 20.4; 26.2 & 21.6; 36.2 & 6.95; 20.1 & 11.4; 15.1 \\
\end{tabular}
\end{threeparttable}
\begin{tablenotes}
\small
\item Comments to Tables \ref{table2a} and \ref{table2b}: Parameters were obtained using following fixed values: $T_2=4120$ K, $R_2=0.144 a_0$, $\xi= 0.8077 a_0$, $a_0$ is the distance between centres of masses of components, $0.5\beta_d$ is the semi-thickness of the outer border of the disk (in degrees), $a$, $e$, $\alpha_e$ are the major semi-axis, the eccentricity, and the azimuth of the periastron of the disk respectively, $R_2$, $\xi$, $a$ are computed during solving the problem; $\chi^2_{crit}=\chi^2_{n-m,0.01}$ is the critical value of $\chi^2$ for the significance level 1\%, $n$, $m$ are the quantity of average dots in the light curve and the number of variables ($m=11$) respectively. $(F_d+F_{HL})/F_{full}$, $F_d/F_{full}$, and $F_{HL}/F_{full}$ are relative contributions of different non-stellar radiating elements to the total system's flux averaged over the orbital period, the disk plus the hot line, the disk only, the hot line only, respectively. Tables \ref{table2a} and \ref{table2b} also contain formal estimation of errors of parameters (except the orbit's inclination $i$) that correspond to the 10\% increase of the $\chi^2$ minimal value.
\end{tablenotes}
\end{table*}

We emphasize that the value of $i$ in our model was found without any observational information about the non-stellar component's luminosity. This luminosity in our case can be found as the solution of the inverse problem of the interpretation of the light curve (see Tables \ref{table2a} and \ref{table2b}) and can be compared with corresponding spectrophotometric estimates. Unfortunately, the scatter of observational estimates of the non-stellar component's contribution to the total luminosity of the system is significant. Besides, it seems, that this contribution is different for different epochs of observations (see Introduction). Using data by \citet{wagner2001,zurita2002,mcclintock2003,torres2004} we can accept an average observational contribution of the non-stellar component to the total flux of the KV UMa system in the optical range ($\lambda=5800\div 6400$\r{A}) in quiescence as 40\%$\div$60\%. In IR in quiescence the observational estimate of the non-stellar contribution according to \citet{mikolajewska2005,khargharia2013} is 30\%$\div$50\%.

The comparison of observational estimates and theoretical models (obtained from the interpretation of light curves) of the non-stellar component's luminosity $(F_d+F_{HL})/F_{full})$ in the optical and IR ranges (see Tables \ref{table2a} and \ref{table2b}) shows that there is a satisfactory agreement between these estimates, and it indicates the reliability of obtained results of modelling. Using our value $i=74^{\circ}\pm 4^{\circ}$ and the mass function of the optical star $f_v(M)=6.1\pm 0.3 M_{\odot}$ the masses of components can be found:

$$
M_{BH}=f_v(M)\left(1+\frac{1}{q}\right)^2\frac{1}{\sin^3 i};
$$

$$
M_v=\frac{M_{BH}}{q}.
$$

Using the orbit's inclination $i=74^{\circ}\pm 4^{\circ}$ for $q=37$ we found: $M_{BH}=7.24^{+0.9}_{-0.7}M_{\odot}$, $M_v= 0.20\pm 0.02M_{\odot}$. For $q=73$ and the same $i$ calculations gave following results: $M_{BH}=7.06^{+0.87}_{-0.69}M_{\odot}$, $M_v=0.10\pm 0.01M_{\odot}$.

These quantities of the orbit's inclination and the black hole mass are close to estimations by \citet{khargharia2013}. The value $i=74^{\circ}\pm 4^{\circ}$ obtained in the present study
is consistent with the value $i=80^{+1}_{-4}$ found by \citet{khruzina2005} within the error limits.

\subsection{Spectrum of non-stellar component}

\begin{figure*}
\renewcommand{\thefigure}{\arabic{figure}a}
\center
\includegraphics[width = \textwidth]{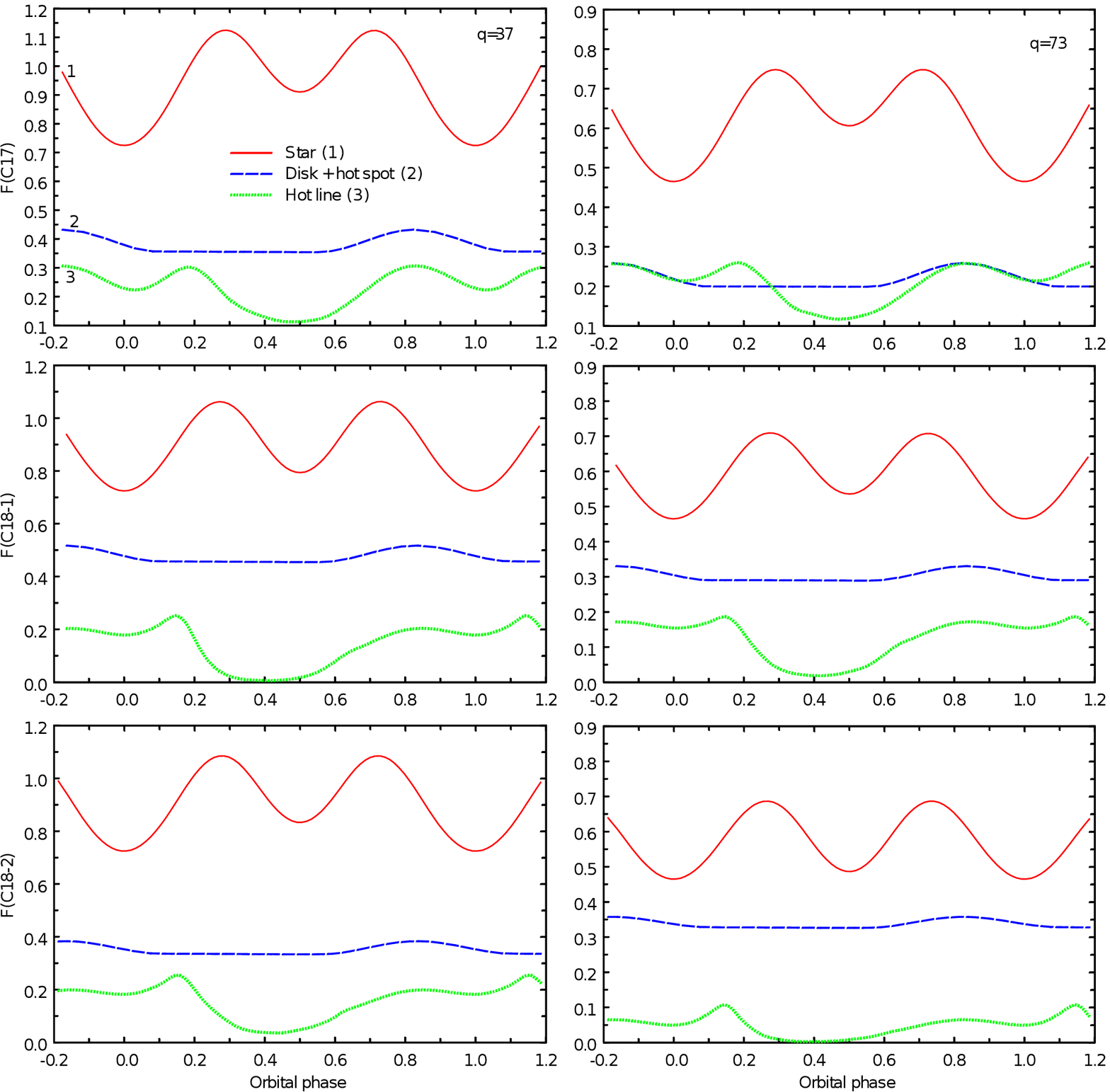}
\vspace{5pt} \caption{Contributions of different components in the total luminosity of the system (in arbitrary units) for optimal parameters of the KV UMa system (see Tables \ref{table2a} and \ref{table2b})  in the integral light in 2017 ($C17$), in 2018 ($C18-1$, $C18-2$).}
\label{figure9a}
\end{figure*}

\begin{figure*}
\addtocounter{figure}{-1}
\renewcommand{\thefigure}{\arabic{figure}b}
\center
\includegraphics[width = \textwidth]{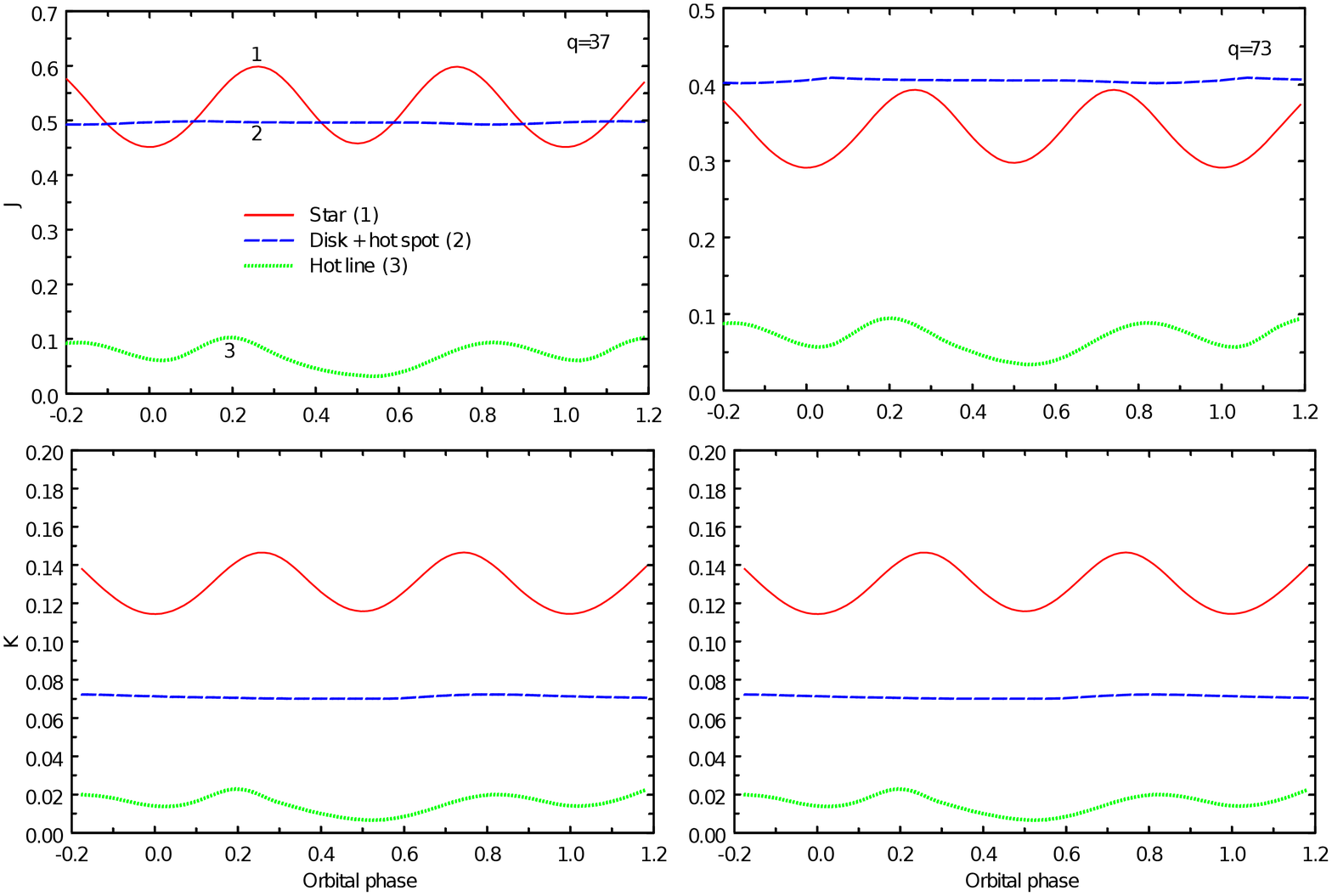}
\vspace{5pt} \caption{The same as Fig. \ref{figure9a} for \textit{J} and \textit{K} bands.}
\label{figure9b}
\end{figure*}

Figs. \ref{figure9a} and \ref{figure9b} shows contributions of different components in the total luminosity of the KV UMa system for optimal model parameters for $i=74^{\circ}$ in the integral light and in bands \textit{J} and \textit{K}, marks are following: ``1'' is the optical star, ``2'' is the accretion disk with the hot spot, ``3'' is the hot line. It is evident that the radiation of the tidally deformed donor star (a symmetric light curve; the star was warmed up with the radiation of the non-stellar component) was summarised with the radiation of the hot line that gave a non-symmetric light curve which contribution changed from one epoch to another. This fact explained the complicated and variable form of observational light curves of KV UMa. The change of the hot line luminosity apparently was related with the change of the mass transfer rate through the point L1 that strongly depends (as $\left(\frac{\Delta R}{R}\right)^3$) on the degree of overflow of the Roche lobe by the donor star which can slightly change due to the chromospheric activity of the optical star \citep{cherepashchuk2019}.

The dependencies of luminosities of different radiating elements averaged over the orbital period  (the disk with the hot spot $F_d$, the hot line $F_{HL}$, and their sum $F_d+F_{HL}$) in units of the total luminosity of the system ($F_{full}$) on the orbit's inclination $i$ for optical and IR light curves are shown in Figs. \ref{figure10a} and \ref{figure10b}. It can be seen that in our model the luminosity of the disk with the hot spot and the luminosity of the hot line change in antiphase as functions of the orbit's inclination, so the total luminosity of the non-stellar component (the disk with the hot spot and the hot line) practically does not change with $i$.

Our IR light curves (see Fig. \ref{figure2b}) in average are $0.23^m$ brighter, bluer, and with a lower amplitude of the orbital variability in comparison with light curves and colours by \citet{mikolajewska2005}. The relative luminosity of the non-stellar component in them turned out to be higher in IR than in the optical range. From Figs. \ref{figure10a} and \ref{figure10b} it is evident that the anomalous increase of the non-stellar component's luminosity in IR originates mostly from the increase of the luminosity of the accretion disk, and the hot line's luminosity does not show strong anomalies.

From the modelling of light curves we know the contribution of each component (the donor star, the disk with the hot spot, the hot line) to the total luminosity of the system (see Figs. \ref{figure9a} and \ref{figure9b}). If we correct the observed average brightness taking into account the interstellar absorption (which in case of KV UMa is very weak) and measure it in absolute energetic units we are able to find the spectrum of each component in the range $\lambda 6400\div 22000$\r{A} in absolute energetic units and restore the full spectrum of the non-stellar component (the disk with the hot spot plus the hot line). Relative contributions of each component averaged over the orbital period are shown in Tables \ref{table2a} and \ref{table2b} as functions of time (only in white colour), and as functions of the wavelength and the orbit's inclination $i$.

\begin{figure*}
\renewcommand{\thefigure}{\arabic{figure}a}
\center
\includegraphics[width = \textwidth]{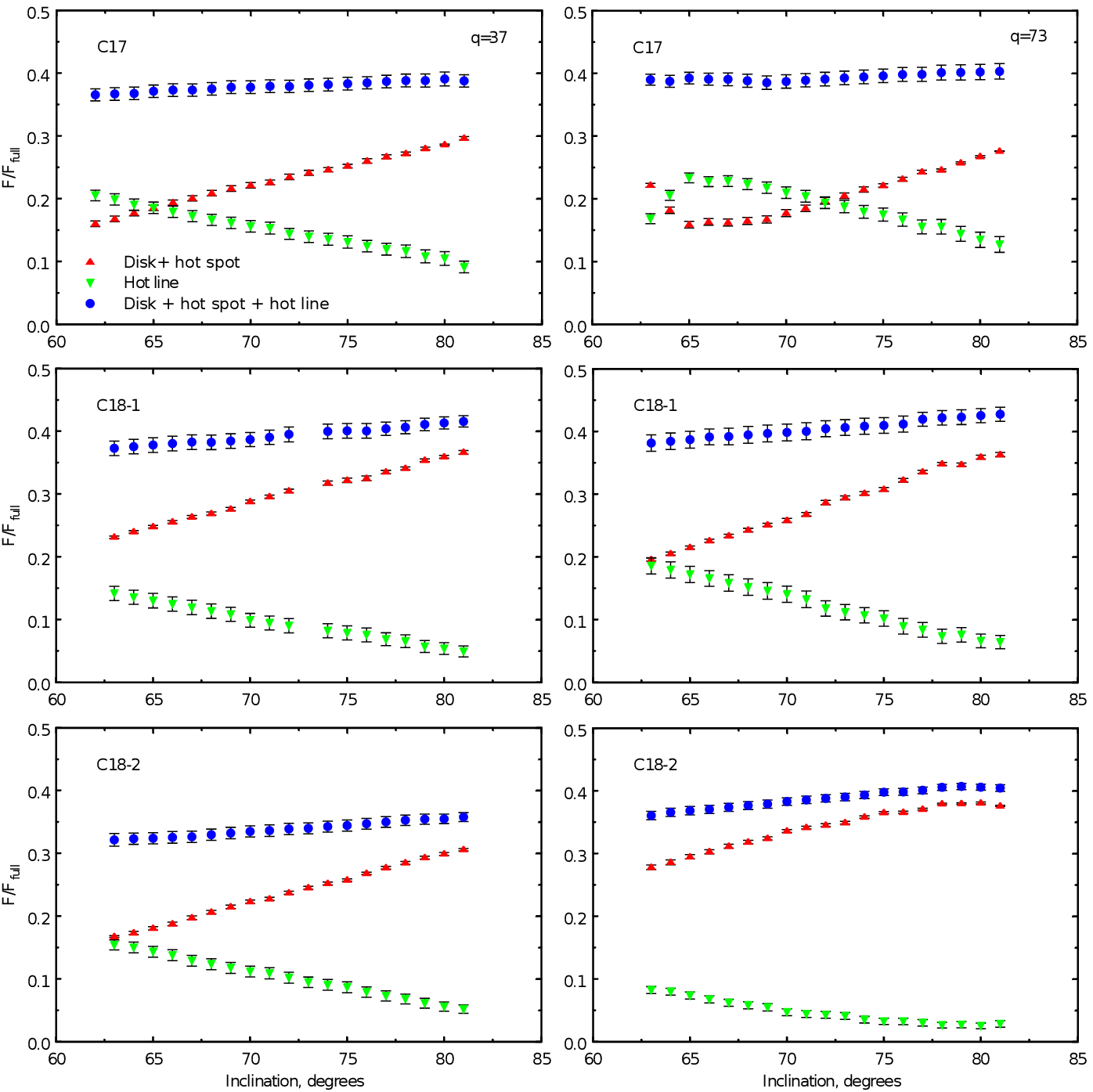}
\vspace{5pt} \caption{Dependencies of contributions (averaged over the orbital period) of different elements on the orbit's inclination $i$ in the integral light ($C17$, $C18-1$, $C18-2$). All units are units in the total system's luminosity $F_{full}$.
}
\label{figure10a}
\end{figure*}

\begin{figure*}
\addtocounter{figure}{-1}
\renewcommand{\thefigure}{\arabic{figure}b}
\center
\includegraphics[width = \textwidth]{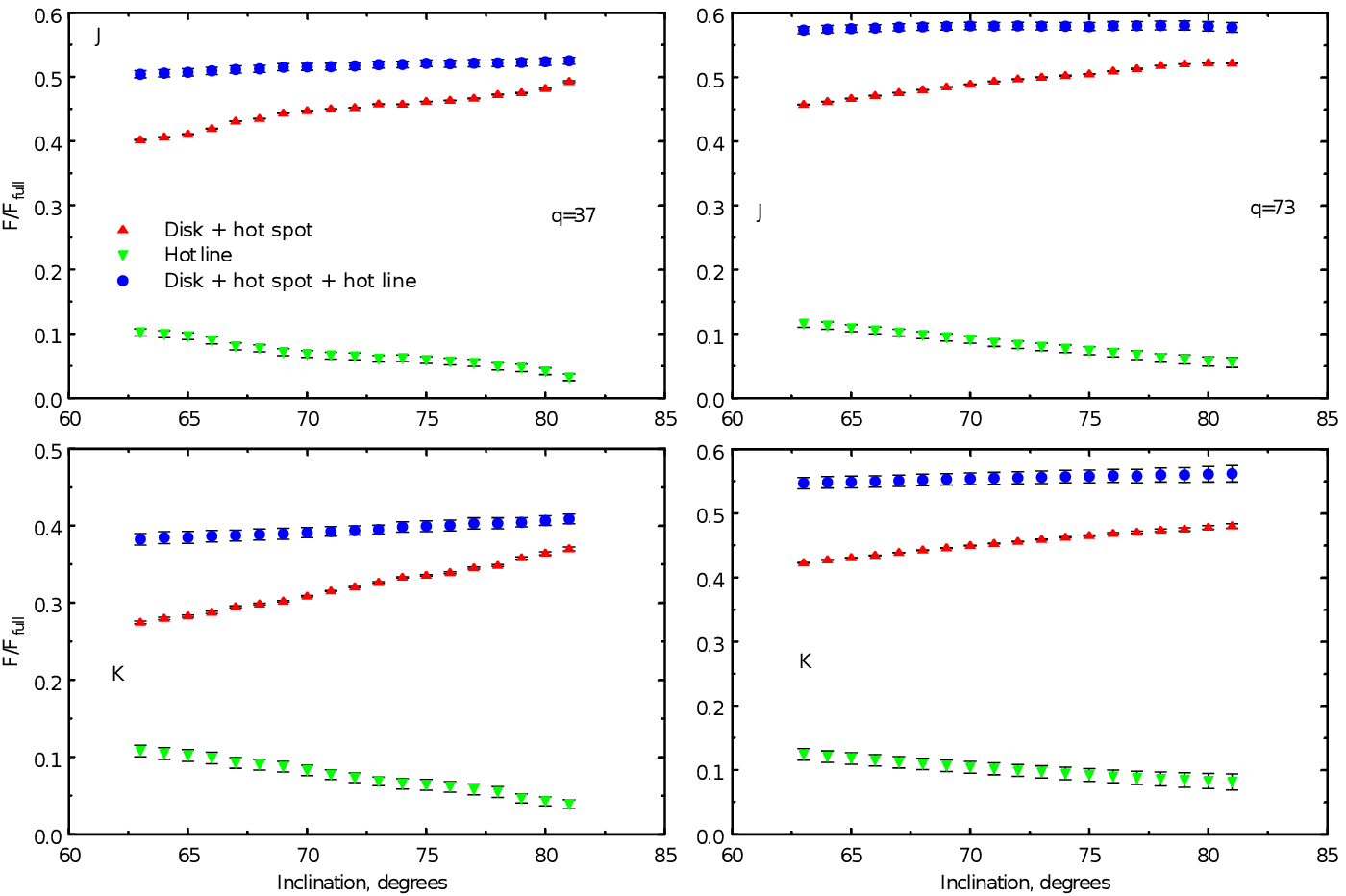}
\vspace{5pt} \caption{The same as Fig. \ref{figure10a} for \textit{J} and \textit{K} bands.
}
\label{figure10b}
\end{figure*}

\begin{table}
\renewcommand{\thetable}{\arabic{table}a}
\large
\centering
\caption{Average stellar magnitudes (IR magnitudes are shown in the MKO photometric system, \citealp{tokunaga2005}).}
\label{table3a}
\begin{tabular}{@{}cc@{}}
\hline
Band & mag \\
\hline
$R(C)$ & 18.93$\pm$0.01 \\
$J$ & 17.79$\pm$0.02 \\
$H$ & 17.12$\pm$0.03 \\
$K$ & 17.06$\pm$0.02 \\
\end{tabular}
\end{table}

\begin{table}
\addtocounter{table}{-1}
\renewcommand{\thetable}{\arabic{table}b}
\large
\centering
\caption{Spectral densities of fluxes, reddened and dereddened in units of $10^{-17}$ erg cm$^{-2}$ s$^{-1}$ \r{A}$^{-1}$ (observed and with accounting for the interstellar reddening).}
\label{table3b}
\begin{tabular}{@{}ccc@{}}
\hline
Flux & reddened & dereddened \\
\hline
$F(R_c)$ & 5.83$\pm$0.02 & 6.01$\pm$0.02 \\
$F(J)$ & 2.31$\pm$0.02 & 2.34$\pm$0.02 \\
$F(H)$ & 1.67$\pm$0.14 & 1.68$\pm$0.14 \\
$F(K)$ & 0.60$\pm$0.01 & 0.60$\pm$0.01 \\
\end{tabular}
\end{table}

Averaged over the orbital period stellar magnitudes of KV UMa in $C(R)$, \textit{J}, and \textit{K} bands (see Table \ref{table3a}) were corrected taking into account the interstellar absorption. According to estimations of different authors the colour excess is small: $E_{B-V}=0.013\div 0.017$ (see, e.g., \citealp{hynes2000,chaty2003}). Assuming an average value $E_{B-V}=0.015$ and following the interstellar absorption law by \citet{rieke1985} we find the interstellar absorption $A_V = 0.046^m$ that coincides with the value $A_V  = 0.05^m$ by \citet{shahbaz2005}. To calibrate the optical observations in the integral light we use the \textit{R} Johnson's filter ($\lambda_{eff}\approx 7000$\r{A}) with the central wavelength that is close to the central wavelength of the \textit{C} band ($\lambda_{eff}\approx 6400$\r{A}). To re-calculate dereddened averaged over the orbital period stellar magnitudes of KV UMa to absolute spectral densities of fluxes $F_{\lambda}$ we used formula

$$
F_{\lambda}=F^0_{\lambda}\cdot 2.512^{(m_o-m)},
$$

\noindent where $m_o=0$, $m$ is the observed stellar magnitude averaged over the orbital period (corrected accounting the interstellar absorption), $F^0_{\lambda}$ is the absolute spectral density of the flux of the zero stellar magnitude star outside the Earth's atmosphere.

For the \textit{R} band we used the calibration by \citet{bessel1998} $F^0_R=2.177\cdot 10^{-9}$, to find IR spectral densities of fluxes we used the calibration by \citet{tokunaga2005} for MKO
$F^0_J=3.01\cdot 10^{-10}$, $F^0_H=1.18\cdot 10^{-10}$,  $F^0_K=4.00\cdot 10^{-11}$ in units of erg cm$^{-2}$ s$^{-1}$ \r{A}$^{-1}$. These IR calibrations are close to data by \citet{koornneef1983,bessel1998}.

Our average values of \textit{J}, \textit{H}, \textit{K} stellar magnitudes were obtained during $\approx 130$ days (from December 2017 to April 2018), also we averaged values in the integral light during 1 year (from November 2017 to November 2018) neglecting average variability about $0.1^m$ during all optical observations. Data about average observed stellar magnitudes and flux densities (reddened and dereddened) are shown in Tables \ref{table3a} and \ref{table3b}. 

\begin{figure}
\center
\includegraphics[width = \columnwidth]{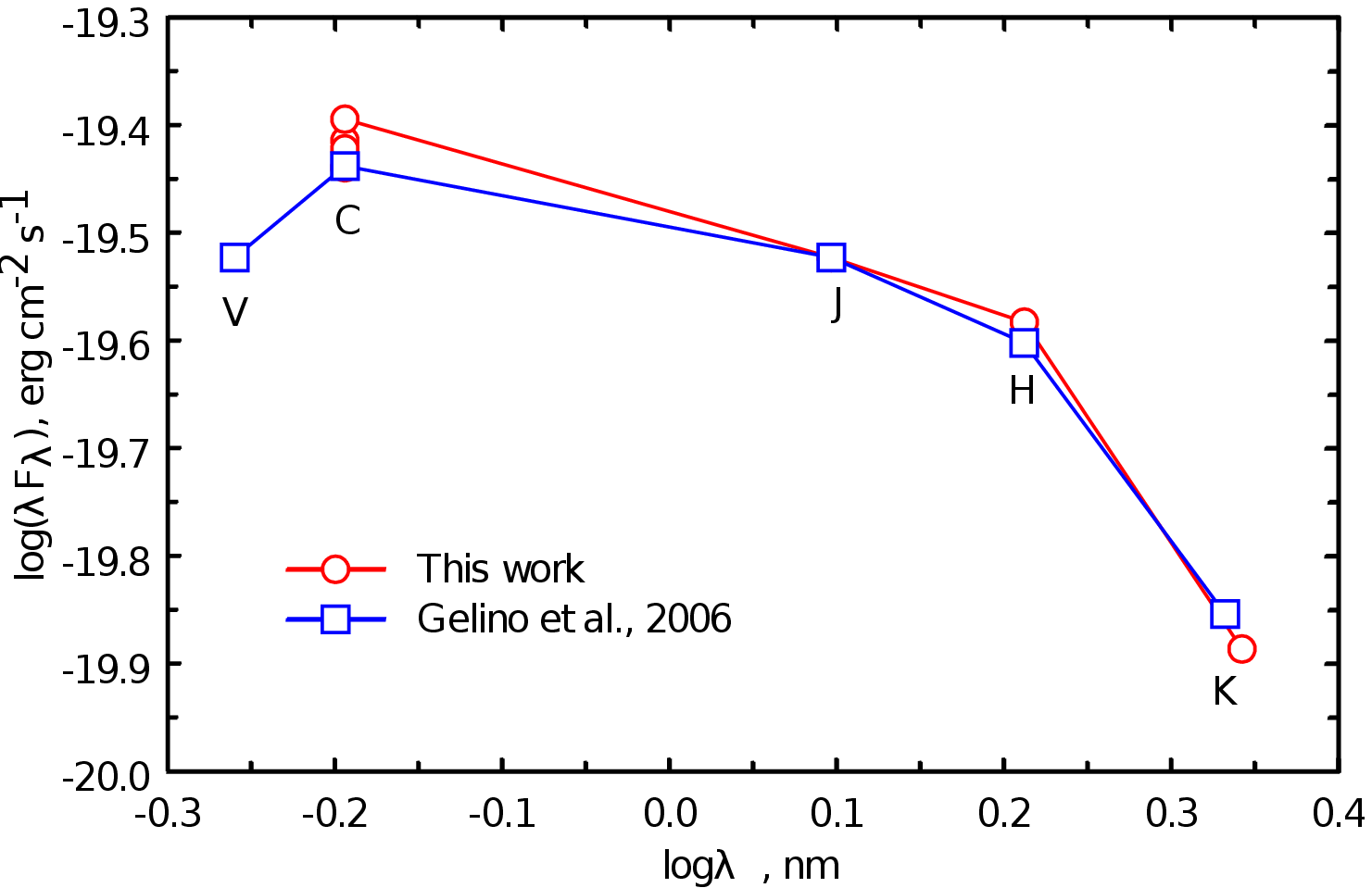}
\vspace{5pt} \caption{A comparison of the observed dereddened energy distribution in the KV UMa spectrum according to our data and according to \citet{gelino2006}, letters depict band names.}
\label{figure11}
\end{figure}

\citet{gelino2006} calculated \textit{BVRJHK} distribution of the energy in the KV UMa spectrum and showed that in \textit{B} and \textit{V} band the system's flux was 65\% and 33\% higher than the flux of a  K7V star, respectively. We conducted a comparison of the red end the spectrum (\textit{RJHK}) assuming that our flux in units $\lambda F(\lambda)$ in \textit{J} filter was equal to the flux by \citet{gelino2006} and realised a good coincidence of them in IR (see Fig. \ref{figure11}). It is important to note that \citet{gelino2006} obtained their data in January 2003. 

\begin{figure*}
\center
\includegraphics[width = 180 mm]{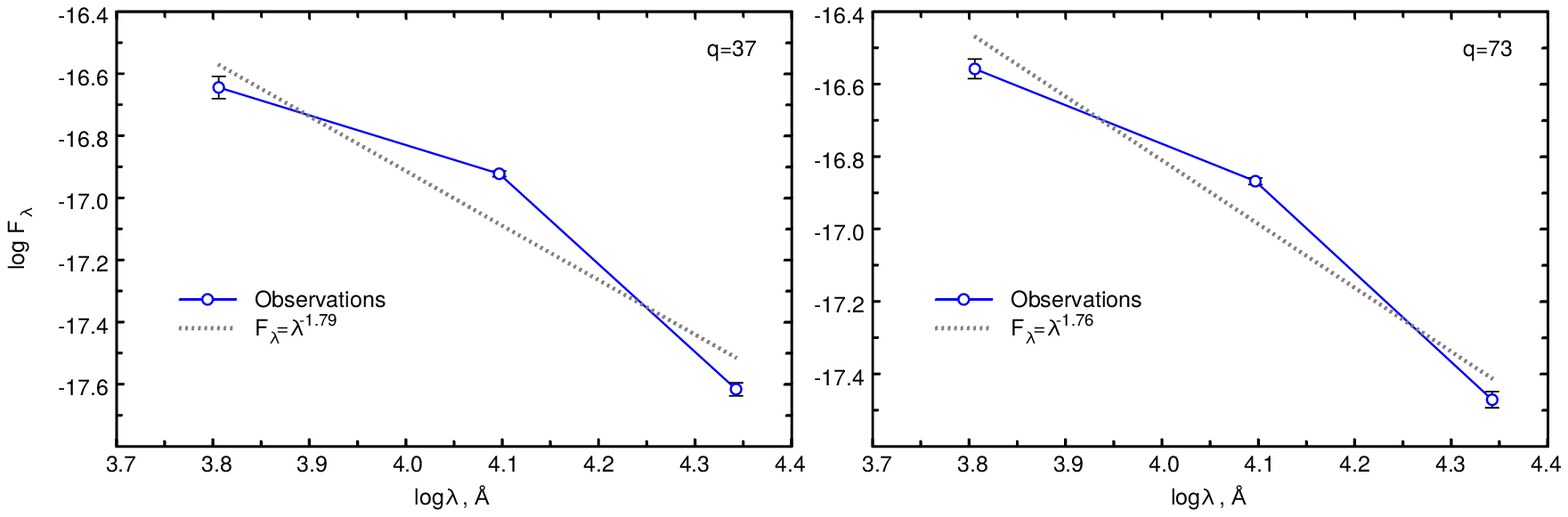}
\vspace{5pt} \caption{Spectra of the non-stellar component in the range $\lambda=6400\div 22000$\r{A} for $q=37$ (left) and $q=73$ (right).}
\label{figure12}
\end{figure*}

\begin{figure*}
\center
\includegraphics[width = 180 mm]{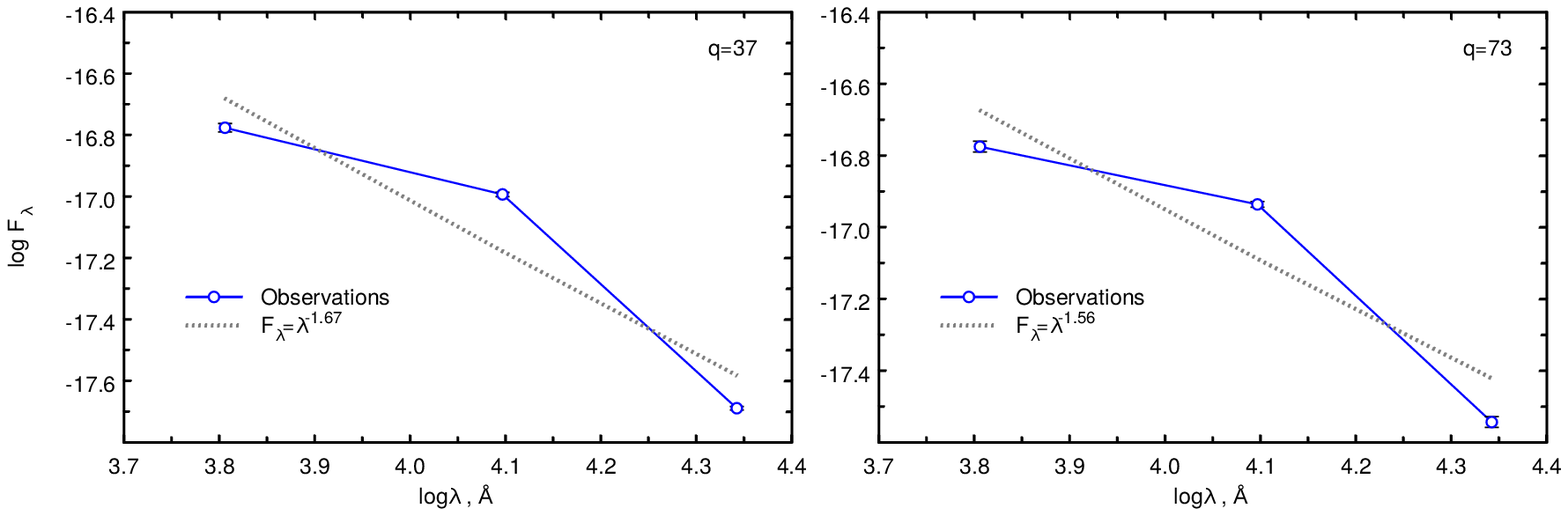}
\vspace{5pt} \caption{Spectra of the radiation of the accretion disk with the hot spot for $q=37$ (left) and $q=73$ (right).}
\label{figure13}
\end{figure*}

In Table \ref{table4} we collected spectral flux densities of KV UMa for $\lambda=6400$, 12500 and 22000\r{A} averaged over the orbital period. The model flux densities are shown: $F_{star}$ for the donor star, $F_d$ for the disk with the hot spot, $F_{HL}$ for the hot line, $F_d+F_{HL}$ for the whole non-stellar component. The observed flux density $F_{\lambda}^{obs}$ is shown taking into account the interstellar absorption.

\begin{table*}
\large
\centering
\caption{Observed dereddened fluxes of KV UMa averaged over the orbital period and model fluxes in units $10^{-17}$ erg cm$^{-2}$ s$^{-1}$ \r{A}$^{-1}$.}
\label{table4}
\begin{tabular}{@{}cccccc@{}}
\hline
Wavelength & $F_{\lambda\ obs}=F_{full}$ & $F_{star}$ & $F_d$ & $F_{HL}$ & $F_d+F_{HL}$ \\
$\lambda$, $\mu m$ & & & & & \\
\hline
\multicolumn{6}{c}{$q=37$} \\
0.64 & $6.014\pm 0.022$ & $3.747\pm 0.126$ & $1.672\pm 0.048$ & $0.590\pm 0.168$ & $2.262\pm 0.013$ \\
1.25 & $2.335\pm 0.024$ & $1.137\pm 0.063$ & $1.016\pm 0.007$ & $0.182\pm 0.014$ & $1.198\pm 0.012$ \\
1.63 & $1.679\pm 0.106$ & -- & -- & -- & -- \\
2.20 & $0.604\pm 0.018$ & $0.266\pm 0.029$ & $0.205\pm 0.006$ & $0.037\pm 0.004$ & $0.242\pm 0.004$ \\
\multicolumn{6}{c}{$q=73$} \\
0.64 & $6.014\pm 0.022$ & $3.620\pm 0.030$ & $1.678\pm 0.054$ & $0.716\pm 0.162$ & $2.394\pm 0.185$ \\
1.25 & $2.335\pm 0.024$ & $0.981\pm 0.024$ & $1.160\pm 0.024$ & $0.194\pm 0.016$ & $1.354\pm 0.028$ \\
1.63 & $1.679\pm 0.106$ & -- & -- & -- & -- \\
2.20 & $0.604\pm 0.018$ & $0.266\pm 0.012$ & $0.287\pm 0.010$ & $0.050\pm 0.008$ & $0.337\pm 0.017$ \\
\end{tabular}
\end{table*}

Fig. \ref{figure12} shows in a logarithmic scale spectra of the non-stellar component $F_d+F_{HL}$ computed from the modelling of light curves in $C(R)$, \textit{J}, \textit{K} bands for two mass ratios $q=37$ and 73. The non-stellar component's spectrum in the range $\lambda=6400\div 22000$\r{A} can be fitted using a power law $F_{\lambda}\sim \lambda^{\alpha}$, where $\alpha=-1.79$ for $q=37$ (left) and $\alpha=-1.76$ for $q=73$ (right).

We also made analogous spectra for the accretion disk with the hot spot $F_d(\lambda)$ excluding the hot line's contribution. Results are shown in Fig. \ref{figure13}. We used a fit $F_d(\lambda)\sim \lambda^{\alpha}$, where $\alpha=-1.67$ for $q=37$ (left) and $\alpha=-1.56$ for $q=73$ (right). These results are interesting to test models of advection dominated accretion flows around black holes in X-ray novae in quiescence (e.g. \citealp{narayan1996,esin1997}).

Powers in the non-stellar component's spectrum $\alpha=-1.79$ and $-1.76$ for KV UMa differ from corresponding powers in the non-stellar component's spectrum of another X-ray nova A0620-00 $\alpha=-2.13$ for a passive state, but close to $\alpha=-1.85$ for an active state \citep{cherepashchuk2019}.

\section{Several generalizations of the model}

We conducted an interpretation of KV UMa light curves using a comparatively simple model that has 11 free parameters. In the frames of this model we successfully gave a reliable determination of the inclination of the system's orbit, refined the masses of the black hole and of the optical star, and restored the non-stellar component's spectrum. Since there are different subtle effects in the system (a chromospheric activity of the donor star, a precession of the accretion disk, see Introduction) it is interesting to consider several generalizations of our model.

\subsection{A model with a spotted optical star}

For our optical light curve $C18-2$ (November 2018, $n=30$, see Fig. \ref{figure8a}) the model of the close binary system in case of $q=37$ was poorly adequate. Therefore to interpret this light curve another model was used, the model of the system with the spotted optical star and with the accretion disk without the region of the interaction between the stream and the disk. As the result the quantity of variables was $m=12$, 8 of them belonged to two ``cold'' spots. In the new model the difference of maxima of the light curve can be explained by the presence of spots on the donor star. A description of the model can be found in papers by \citet{khruzina1995,cherepashchuk2019}.  

The minimal residual in this case ($\chi^2_{min}=55.72$) lied higher than the critical value $\chi^2=34.8$ within the 1\% significance level, i.e., the spotted model along with the hot line model should be rejected by observations. Nevertheless, even in the frames of this not fully adequate model, we independently found a reasonable estimate of the orbit's inclination $i=73^{\circ}$, which almost does not depend on the definite model of the system.

\subsection{A model with a precessing accretion disk}

According to spectroscopic data \citep{shahbaz2004,zurita2016} in the KV UMa system there are long periodical shifts of the centroid of the $H_{\alpha}$ emission line with a period of 52 days, which were interpreted as the rotation of the semi-major axis of the elliptical disk (precession) with the period which coincides with the period of superhumps on the light curves during the flash attenuation (see, e.g., \citealp{zurita2002}). Since in our model the disk was elliptical, we tried to find from our optical light curves (that cover the time interval of 350 days) the change of the disk's orientation with time. For the search we used three optical light curves of KV UMa obtained in November 2017, in June 2018, and in November 2018. The model with the hot line and with the elliptical accretion disk was used. The disk's orientation $\alpha_e$ and its eccentricity $е$ are shown in Tables \ref{table2a} and \ref{table2b} and are given for clarity in Table \ref{table5}. The results depend on $q$ weakly. The disk's eccentricity $e=0.019\div 0.025$ practically did not change with time. The disk's orientation angle $\alpha_e$ that characterized the disk's precession was in average $125^{\circ}$ in November 2017, $90^{\circ}\div 95^{\circ}$ in June 2018, and $145^{\circ}\div 146^{\circ}$ in November 2018.

If one assumed \citep{shahbaz2004,zurita2016} that $P_{prec}\approx 50\div 60$ days, in case of the clockwise disk's rotation, i.e., the retrograde rotation according to the orbital rotation, the disk's periastron had in average 6.943 cycles during $\approx 348$ days (341-354) from November 2017 to November 2018, $P_{prec}\approx 50.1(9)$ days. If the rotation of the disk was counter clockwise, i.e. it had the same direction with the orbital rotation, the disk made 7.057 cycles, $P_{prec}\approx 49.3(9)$ days. Within errors of $P_{prec}$ values for the clockwise and counter clockwise rotation were equal, so longer observational sets are needed to clarify this issue.

\begin{table*}
\large
\centering
\caption{The eccentricity of the accretion disk and the angle of its orientation as functions of time.}
\label{table5}
\begin{tabular}{@{}cccc@{}}
\hline
Parameters & $C17$ & $C18-1$ & $C18-2$ \\
JD 245 0000+ & 8082-8083 & 8271-8279 & 8424-8436 \\
\hline
\multicolumn{4}{c}{$q=37$} \\
$e$ & $0.025\pm 0.004$ & $0.019\pm 0.004$ & $0.022\pm 0.003$ \\
$\alpha_{е}$, $^{\circ}$ & $125\pm 2$ & $95\pm 2$ & $145\pm 15$ \\
\multicolumn{4}{c}{$q=73$} \\
$e$ & $0.025\pm 0.001$ & $0.021\pm 0.004$ & $0.022\pm 0.004$ \\
$\alpha_{е}$, $^{\circ}$ & $125\pm 2$ & $90\pm 18$ & $146\pm 5$ \\
\end{tabular}
\end{table*}

As follows from Table \ref{table5} the orientation angle of the disk changes from one epoch to another, so it can reflect the precession of the disk. To reliably find the disk's precession from light curves it is necessary to obtain a dense number of observations during several months. We suppose to make it in future.

\section{Conclusions}

We obtained optical and IR observations of KV UMa in its quiescent state in 2017-2018 and modelled corresponding light curves. Here we summarize main results obtained in these studies.

\begin{enumerate}[label*=\arabic*.]

\item Our observations in November 2017, June 2018, and November 2018 reveal no transition of the system from a passive to an active states. The system was in the passive state (using terminology by \citealp{cantrell2008}) with a relatively low flickering, although the average brightness of the system in the optical range during 350 days was monotonically decreasing, and in November 2018 turned out to be $\approx 0.1^m$ less than in the beginning of observations.

\item In comparison to the \textit{J} light curve obtained in 2003 and 2004 by \citep{mikolajewska2005}, our \textit{J} light curve obtained in end of 2017 and in 2018 has the average brightness $J=17.79^m\pm 0.02^m$ and the colour index $J-K\approx 0.73^m\pm 0.04^m$, i.e, in IR in our case KV UMa became brighter by $0.23^m$ and bluer, at the same time the amplitude of variations in the \textit{J} light curve dropped from $0.35^m$ to $0.23^m$. These differences apparently are connected with the variability of the contribution of the non-stellar component: the accretion disk with the interaction region. Although the average brightness of the system was changed within wide limits, the flickering in KV UMa light curves was not detected. So, in contrast to the A0620-00 system, where the increase of the average brightness of the system is accompanied with the strong increase of the flickering, in KV UMa the flickering did not appear with the increase of the average brightness of the system.

\item We modelled optical and IR light curves of KV UMa in the frames of the model of interacting binary system that contains the donor star filling its Roche lobe and the accretion disk with the complicated interaction region: the hot line and the hot spot. We justified the adequacy of the model with observational data.

\item The reliable value of the inclination of the system's orbit was found ($i=74^{\circ}\pm 4^{\circ}$) using the modelling of five independent light curves of KV UMa (three in the optical range and two in IR), this value is consistent with the value by \citep{khargharia2013}. Our value of $i$ was used to find mass of the black hole and of the donor star for two mass ratios of components: $M_{BH}=7.24^{+0.9}_{-0.7}M_{\odot}$, $M_v=0.20\pm 0.02M_{\odot}$ for $q=37$, $M_{BH}= 7.06^{+0.87}_{-0.69} M_{\odot}$, $M_v=0.10\pm 0.01 M_{\odot}$ for $q=73$. Due to the fact that $q=73$ was found in the frames of more developed model of the rotational broadening of profiles of absorption lines \citep{petrov2017} the masses $M_{BH}$ and $M_v$ corresponding to $q=73$ are preferable.
     
\item From the modelling of optical and IR light curves we restored the non-stellar component spectrum (the accretion disk plus the interaction region) in the range $\lambda=6400\div 22000$\r{A}, and also the spectrum of the accretion disk (Figs. \ref{figure12} and \ref{figure13}). The non-stellar component's spectrum was satisfactory described by a power law $F_{\lambda}\sim \lambda^{\alpha}$, where $\alpha=-1.79$ for $q=37$ and $\alpha=-1.76$ for $q=73$. These values of $\alpha$ are close to values obtained for the A0620-00 X-ray nova in the active stage: $\alpha=-1.85$ \citep{cherepashchuk2019}. The spectrum of the accretion disk can be fitted in the range $\lambda = 6400\div 22000$\r{A} by a power law $F_{\lambda}\sim \lambda^{\alpha}$, where $\alpha=-1.67$ (for $q=37$) and $\alpha=-1.56$ (for $q=73$), see Figs. \ref{figure12} and \ref{figure13}. These data are interesting for tests of models of advection dominated accretion flows around black holes.

\item We also considered two generalizations of our model and made the interpretation of optical light curves. The model with ``cold'' spots on the optical star did not allow to significantly increase the coincidence between observed and theoretical light curves, and one of spots was near the L1 point potentially blocking the mass transfer between components. Therefore the model with spots on the star was not attractive. For three sets of optical observations of KV UMa (November 2017, June 2018, and November 2018) we attempted to find traces of the precession of the elliptical accretion disk with the period 52 days found using spectrophotometric observations \citep{shahbaz2004,zurita2016}. The change of the disk's orientation angle from one epoch to another was detected indicating the possibility to reveal the precession of the disk from light curves. To solve this problem it is necessary to get a dense number of light curves of KV UMa during several moths, it is planned for the future.

\item Evolutionary aspects of the problem of low mass X-ray binary systems with anomalously rapid decrease of the orbital period are considered in Appendix. It is shown that rapid decrease of the KV UMa orbital period is consistent with the model of the increased magnetic fields in the low mass optical star during the preceding common envelope.

\end{enumerate}

\section*{Acknowledgements}

We thank K. Atapin for his help with observations.

The work of A.~M. Cherepashchuk (interpretation) was supported by the Russian Science Foundation grant \mbox{17-12-01241}, the work of N.~A. Katysheva, T.~S. Khruzina, S.~Yu. Shugarov, A.~M. Tatarnikov (observations and modelling) was supported by the the Program of development of M.~V. Lomonosov Moscow State University ``Leading scientific schools'', project ``Physics of stars, relativistic objects and galaxies''.

This research made use of the equipment (IR camera) purchased from funds of the Program of development of M.~V. Lomonosov Moscow State University.

S. Yu. Shugarov thanks the APVV-15-0458 and VEGA 2/0008/17 grants of Slovak academy of Sciences for partial support.

The authors are grateful to the anonymous referee of the manuscript for valuable comments that helped to improve it.



\bibliographystyle{mnras}
\bibliography{cherepashchuk-kv-uma}


\section*{Appendix: Evolutionary aspects}

KV UMa has an anomalous height over the galactic plane $z=1.7$ kpc and shows an anomalously fast decrease of the orbital period $dP/dt=-1.83\pm 0.66$ ms yr$^{-1}$ that corresponds to the change $-0.85\pm 0.30$ $\mu$s during one orbital cycle. This is about 150 times faster than expected from the radiation of gravitational waves by the system \citep{g-h-2012}. \citet{g-h-2012} suggested the model of the donor star with very strong magnetic field ($> 10-20$ kGs) to explain so fast decrease of the orbital period. In this case the decrease can be explained by the orbital angular momentum loss from the system by the magnetic stellar wind. The authors suggest a fast evolution of the system and a short life time for it. Also there is a model of an X-ray system which anomalous fast decay of the orbital period can be explained by the interaction of the system with the circumbinary envelope \citep{xu2018}.

It is interesting to explore the evolutionary scenario for such X-ray binaries with the ``Scenario Machine'' (it is a computer program for investigations of the evolution of close binary stars with the population synthesis method; \citealp{kornilov1983}) taking into account their short evolution time.

\subsection{The Scenario Machine}

To study the evolution of close binary black holes with low mass non-degenerate companions in this work the ``Scenario Machine'' was applied. With its help it is possible to study statistical properties of a population of stars as well as separate evolutionary tracks of close binary systems. The code was described in detail by \citet{lipunov1996,lipunov2009}, therefore here we described only the most important free evolutionary parameters for systems under investigation.

As the initial mass distribution function of binaries we used the Salpeter function and accepted the equiprobable initial mass ratio of components $Q=M_2/M_1<1$, the initial semi-major axis had a flat distribution in a logarithmic scale in the range from $10R_{\odot}$ to $10^6R_{\odot}$. It also should be noted that the equiprobable distribution on the initial mass ratio in the system for $Q<0.1$ potentially can be too rough assumption due to selection effects \citep{tutukov1988}.

The rate of the mass loss in the stellar wind $\dot M$ is essential. It can change the distance between components as well as the mass of the pre-supernova star. Also it is able to define whether the system can form the common envelope. In case of the loss of the hydrogen envelope by the progenitor of the black hole before the Roche lobe filling the common envelope cannot form, and the X-ray nova does not form too. For main sequence stars and supergiants we used following formula for the mass loss:

\begin{equation}
\dot M=\frac{\alpha L}{cV_{\infty}}
\label{awind}
\end{equation}

\noindent where $L$ is the luminosity of the star, $V_{\infty}$ is the wind's velocity at the infinity, $c$ is the speed of light, $\alpha$ is a free parameter. It is assumed that the full mass loss $\Delta M$ does not exceed 10\% of the hydrogen envelope during the life time in main sequence and supergiant stages. The mass loss by Wolf-Rayet stars was

\begin{equation}
\Delta M_{WR}=\alpha_{WR}M_{WR},
\label{awindwr}
\end{equation}

\noindent where $M_{WR}$ is the initial mass of the Wolf-Rayet star.

The common envelope arises if the Roche lobe is filled by the star with a highly evolved core independently on the mass ratio in the system, in other cases the common envelope forms if the condition $Q=M_2/M_1\le q_{cr} = 0.3$ is met, if $Q>0.3$ the evolution goes without the common envelope (see similar conditions in recent papers by \citealp{heuvel2017,pavlovskii2017}). For systems under investigation this condition is always met. The common envelope stage preceding the supernova explosion is necessary, because the current distances between components of mentioned systems are comparable to the size of a main sequence star that can end its life as a black hole. The common envelope is described by the effectiveness $\alpha_{CE}=\Delta E_b/\Delta E_{orb}$ (where $\Delta E_b$ is the change of the bound energy of the envelope, $\Delta E_{orb}$ is the change in the system's orbital energy), it is computed using formula:

\begin{equation}
\alpha_{CE}\left(\frac{GM_a M_c}{2a_f}-\frac{GM_a M_d}{2a_i}\right)=\frac{GM_d (M_d - M_c)}{R_d},
\label{ace}
\end{equation}

\noindent where $M_a$ is the mass of the accreting star, $M_d$ is the donor star's mass, $M_c$ is the mass of the core of the donor star, $a_i$ is the semi-major axis of the system in the beginning of the common envelope stage, $a_f$ is the semi-major axis in the end of it, $R_d$ is the donor star's radius.

The rate of the angular momentum loss $J$ under the influence of the magnetic stellar wind is determined by equation \citep{tutukov1988}:

\begin{equation}
\label{msw-momentum}
\frac{d\ln J}{dt}=-10^{-14}\frac{R_2^4(M_1+M_2)^2R_{\odot}}{\lambda_{MSW}^2a^5M_1 M_{\odot}} \textrm{s}^{-1},
\end{equation}

\noindent where $R_2$ and $M_2$ are the radius and the mass of the star with the magnetic stellar wind, $M_1$ is the mass of another star, $a$ is the semi-major axis, $\lambda_{MSW}$ is the parameter of the magnetic stellar wind, which value usually is accepted to be equal to 1, in the present study it serves as a free parameter. From Equation \ref{msw-momentum} follows the time scale of the angular momentum loss by the magnetic stellar wind \citep{tutukov2018}:

\begin{equation}
\label{msw-time}
\tau=\frac{3.3a^5M_1\lambda_{MSW}^2}{(M_1+M_2)^2 R_2^4},
\end{equation}

\noindent in this Equation masses are in $M_{\odot}$, the semi-major axis and the radius are in $R_{\odot}$.

\citet{ohlmann2016} conducted magnetohydrodynamical calculations of the common envelope dynamics in the system consisting of a $1M_{\odot}$ main sequence star and a $2M_{\odot}$ red giant. A common envelope in such system may result in the increase of the magnetic field of the main sequence star up to 10-100 kG even after 120 days from the beginning the common envelope stage. The increase of the magnetic field they connected with the magneto-rotational instability. Despite of that the masses considered by \citet{ohlmann2016} are significantly lower than masses of progenitors of black holes, one can accept that in a more or less similar way the accreting star's magnetic field can grow during the common envelope stage.

In the moment of the formation of the black hole a part of the preceding star collapses under the event horizon, and a part of it can be ejected. The black hole mass $M_{BH}$ is calculated as

\begin{equation}
\label{kbh}
M_{BH}=k_{BH}M_{preSN}
\end{equation}

\noindent where $M_{preSN}$ is the pre-supernova mass, $0\le k_{BH}\le 1$ is a free parameter (the part of the pre-supernova star mass that falls under the event horizon). The parameter $k_{BH}$ has an important value, because, firstly, it defines the mass of the forming black hole, and, secondly, if the mass loss during the explosion exceeds 50\% of the total mass of stars before the supernova explosion, the system loses the gravitational bound and decays.

Also in the explosion the star's compact remnant (a black hole) can get an additional velocity

\begin{equation}
v_{BH}=v_a\frac{M_{preSN}-M_{BH}}{M_{BH}},
\label{kickbh}
\end{equation}

\noindent were $v_{BH}$ is the additional black hole's velocity, $v_a$ is distributed as 

\begin{equation}
f(v_a) \sim \frac{v_{a}^2}{v_0^3} e^{-\frac{v_{a}^2}{v_0^2}},
\label{vanis}
\end{equation}

\noindent wher $v_0$ is a free parameter. The velocity's direction is equiprobable. The additional kick is important, because it get lead to the decay of binary system, and in very particular cases it can bound systems (that should decay without it).

\subsection{A determination of value areas for evolutionary parameters}

\begin{table*}
\large
\centering
\caption{An evolutionary track that leads to the formation of a close binary system consisting of a black hole and a low mass non-degenerate star. The magnetic stellar wind is weak ($\lambda_{MSW}=1$). Columns in the table depict following parameters: $System$ is the composition of the binary, $\Delta T$ is the duration of the stage, $M_1$ is the mass of the initially more massive star, $M_2$ is the companion's mass ($M_{1,2}$ in $M_{\odot}$), $a$ is the semi-major axis (in $R_{\odot}$),  $P_{orb}$ is the orbital period in days, $e$ is the eccentricity, $T$ is the time since the beginning of the evolution ($\Delta T$, $T$ in millions of years). Stages are marked as following: ``I'' is the main sequence star, ``II'' is the supergiant, ``3'' is the star filling its Roche lobe, ``WR'' is the Wolf-Rayet star, ``BH'' is the black hole, ``CE'' is the common envelope. Values of evolutionary parameters are: $k_{BH}=0.8$, $\alpha_{CE}=1.0$, $\alpha=\alpha_{WR}=0.3$.}
\label{track1}
\begin{tabular}{@{}cccccccc@{}}
\hline
System & $\Delta T$ & $M_1$ & $M_2$ & $a$ & $P_{orb}$ & $e$ & $T$ \\
\hline
I+I & 4.8 & 31.55 & 0.75 & 750 & 419 & 0 & 0 \\
I+I & & 28.25 & 0.75 & 830 & 515 & 0 & 4.8 \\
II+I & 0.48 & 28.25 & 0.75 & 830 & 515 & 0 & 4.8 \\
II+I & & 19.21 & 0.75 & 1200 & 1079 & 0 & 5.3 \\
3+I, CE & 0.01 & 19.21 & 0.75 & 1200 & 1079 & 0 & 5.3 \\
3+I, CE & & 12.55 & 0.75 & 28 & 4.7 & 0 & 5.3 \\
WR+I & 0.45 & 12.55 & 0.75 & 28 & 4.7 & 0 & 5.3 \\
WR+I & & 8.78 & 0.75 & 39 & 9.15 & 0 & 5.8 \\
\multicolumn{6}{c}{SN Ib} \\ 
BH+I & 1.5E+04 & 7.03 & 0.75 & 50 & 14.7 & 0.23 & 5.8 \\
BH+I & & 7.03 & 0.73 & 49 & 14.28 & 0.19 & $1.5\cdot 10^4$ \\
\end{tabular}
\end{table*}

\begin{table*}
\large
\centering
\caption{The same track as in Table \ref{track1}, the magnetic stellar wind is strong ($\lambda_{MSW}=0.13$). Values of evolutionary parameters (except $\lambda_{MSW}$), values in columns, and marks of evolutionary stages are the same as in Table \ref{track1}, ``MSW'' is the stage where the evolution is strongly affected by the magnetic stellar wind, ``BB'' is the Wolf-Rayet star filling its Roche lobe.}
\label{track2}
\begin{tabular}{@{}ccccccccc@{}}
\hline
System & $\Delta T$ & $M_1$ & $M_2$ & $a$ & $P_{orb}$ & $e$ & $T$ \\
\hline
I+I & 4.8 & 31.55 & 0.75 & 750 & 419 & 0 & 0 \\
I+I & & 28.25 & 0.75 & 830 & 515 & 0 & 4.8 \\
II+I & 0.48 & 28.25 & 0.75 & 830 & 515 & 0 & 4.8 \\
II+I & & 19.21 & 0.75 & 1200 & 1079 & 0 & 5.3 \\
3+I, CE & 0.01 & 19.21 & 0.75 & 1200 & 1079 & 0 & 5.3 \\
3+I, CE & & 12.55 & 0.75 & 28 & 4.7 & 0 & 5.3 \\
WR+I, MSW & 0.2 & 12.55 & 0.75 & 28 & 4.7 & 0 & 5.3 \\
WR+I, MSW &  & 10.82 & 0.75 & 32 & 6.2 & 0 & 5.5 \\
BB+3, MSW, CE & 0.01 & 10.82 & 0.75 & 32 & 6.2 & 0 & 5.5 \\
BB+3, MSW, CE & & 7.90 & 0.34 & 1.7 & 0.09 & 0 & 5.5 \\
\multicolumn{6}{c}{SN Ib} \\ 
BH+3, MSW & 30.65 & 6.32 & 0.34 & 2.2 & 0.15 & 0.24 & 5.5 \\
\multicolumn{5}{c}{MSW stage stops or the optical star decays} & & & 36.15 \\
\end{tabular}
\end{table*}

To find the range of valid values of parameters $\alpha_{CE}$, $\alpha$, $\alpha_{WR}$, $k_{BH}$,  $v_0$ we calculated the quantity of binary systems in the Galaxy taking into account their life times in corresponding stages (assuming that all stars are binary). As the X-ray novae we treated a black hole in pair with a low mass non-degenerate star (with mass from $0.2M_{\odot}$ to $1.1M_{\odot}$) with the orbital period $\lesssim 0.5$ days. Along with it we calculated the quantity of systems with the orbital period less than 1.5 days in order to show that the common envelope and the flat initial distribution of systems on the semi-major axis in a logarithmic scale give the enough quantity of binaries with periods longer than periods of known X-ray novae, so if the magnetic stellar wind strength grows the quantity of X-ray novae remains adequate to observational quantities.

According to our calculations the non-zero additional velocity of the black hole acquired during its formation $v_0\leq 10$ km s$^{-1}$ practically does not change the quantity of the systems under investigation, the increase of $v_0$ over 10 km s$^{-1}$ leads to a very rapid decrease of their quantities down to zero for $v_0\approx 100$ km s$^{-1}$. Therefore the presence of X-ray novae in the Galaxy and estimations of their quantities taking into account selection effects allows to conclude that black holes in such systems apparently do not acquire a significant kick during their formation. For this reason we use the kick velocity value $v_0=0$. The high distance of KV UMa from the galactic plane probably can be explained by another process (e.g., the dynamical interaction with another body) or by very specific direction of the kick and adjusted value of it.

The rate of mass loss via the stellar wind was varied from low ($\alpha=\alpha_{WR}=0.1$) to high ($\alpha=\alpha_{WR}=0.7$). As follows from calculations, the quantity of studied systems in the Galaxy depends on the strength of the wind weakly if $\alpha\lesssim 0.5$, $\alpha_{WR}\lesssim 0.5$, if the wind's strength grows further the quantity of X-ray novae rapidly falls to zero. For calculations below we took the value $\alpha=\alpha_{WR}=0.3$ (it approximately corresponds to the solar metallicity), which allows to make a good estimation of the quantity of systems under consideration.

\begin{figure}
\center
\includegraphics[width=\columnwidth]{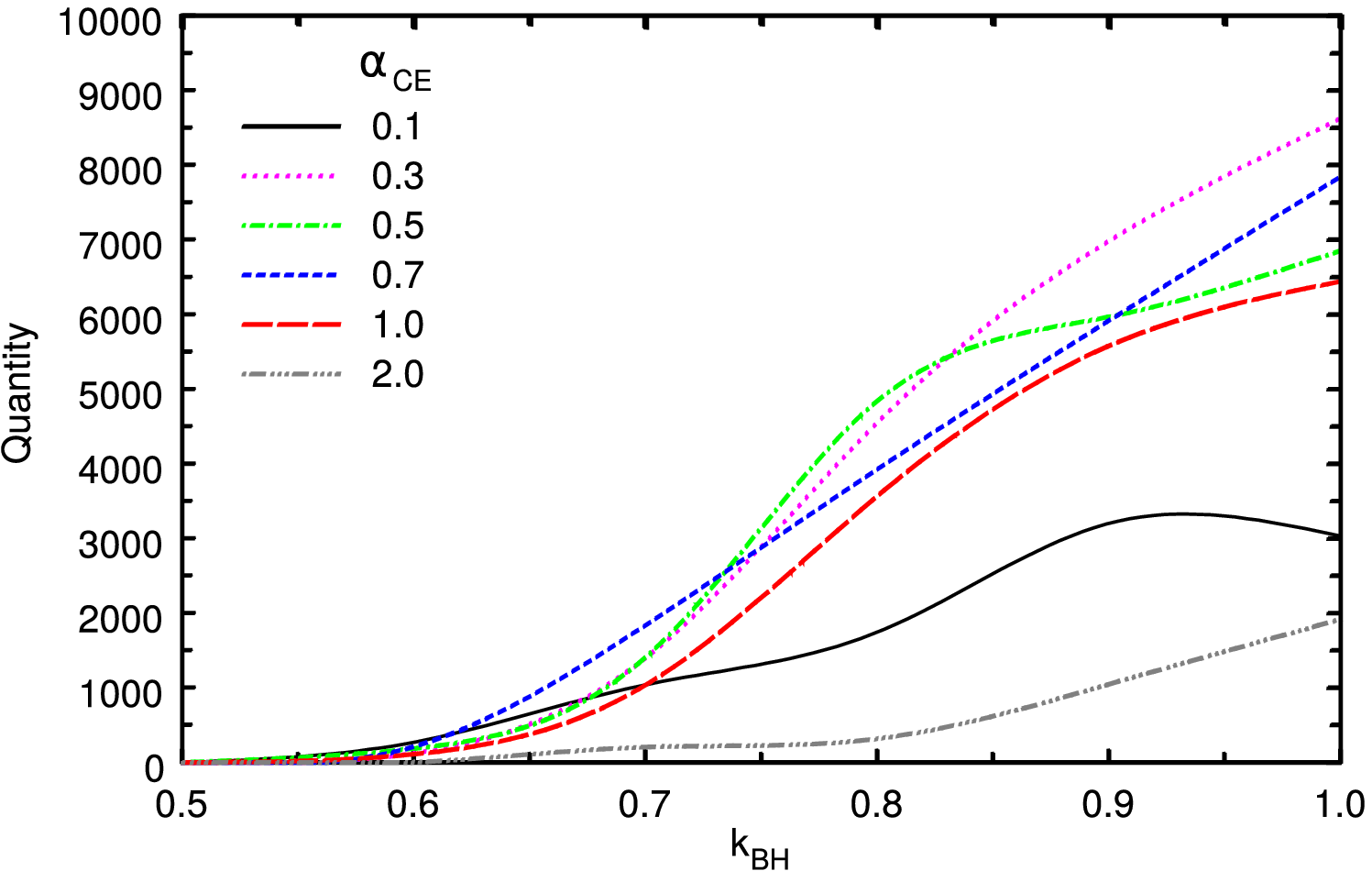}
\vspace{5pt} \caption{The number of X-ray novae in the Galaxy that consist of black holes and main sequence stars (including those that fill their Roche lobes on the main sequence) with masses between $0.2R_{\odot}$ and $1.1R_{\odot}$. The orbital period is $P_{orb}\le 0.5$ days. Following evolutionary parameters were used: $\alpha=\alpha_{WR}=0.3$, $\lambda_{MSW}=1$, $v_0=0$.}\label{f1}
\end{figure}

Figs. \ref{f1} and \ref{f2} show the quantity of studied systems calculated using different sets of evolutionary parameters, the value of $\lambda_{MSW}$ was assumed to be 1. According to observational estimates the quantity of X-ray novae in the Milky Way is approximately $300\div 3000$ (according to \citealp{chen1997} there are at least 100 X-ray novae with black holes, an average density of such systems is 0.25 per square kpc, it gives a rough estimation of three thousand system in the Milky Way). A modern account for selection effects allows to only approximately estimate lower and upper limits of the quantity of black holes with low mass non-degenerate stars \citep{arur2018}.
This is very soft limit that does not allow to fix parameters. To met the mentioned condition we chose following set of evolutionary parameters for calculations with the modification of the strength of the magnetic stellar wind: $k_{BH}=0.8$, $\alpha_{CE}=1.0$, $\alpha=\alpha_{WR}=0.3$ (evolutionary tracks in Tables \ref{track1}, \ref{track2}, and in Fig. \ref{f3} were calculated with this set too). This set allows to get an adequate quantity of studied binaries. Also it should be noted that in spite of the fact that low mass X-ray binaries mainly were formed in early stages of the evolution of the Galaxy (see Fig. 2 by \citealp{yungelson2006}), our conclusions does not change due to very long life times of these binaries, because the ``stockpile'' of binaries in wider orbits is enough to supply the observed quantity even if there are physical reasons for more rapid approach of components than it was expected from gravitational wave radiation angular momentum losses.

\subsection{Calculations with the variation of the magnetic stellar wind strength}

\begin{figure}
\center
\includegraphics[width=\columnwidth]{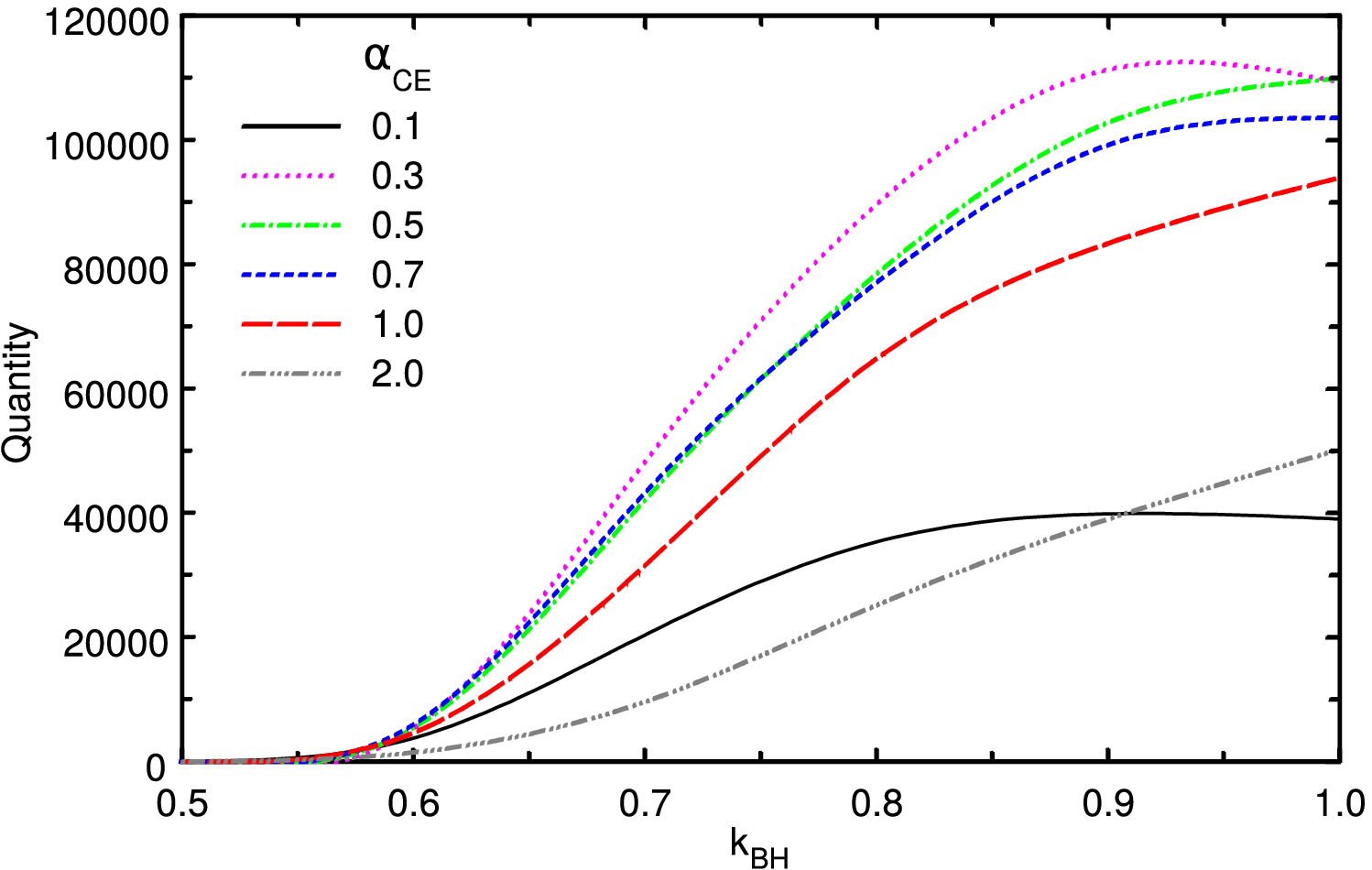}
\vspace{5pt} \caption{The same as Fig. \ref{f1}, the orbital period is $P_{orb}\le 1.5$ days.}\label{f2}
\end{figure}

\begin{figure}
\center
\includegraphics[width=\columnwidth]{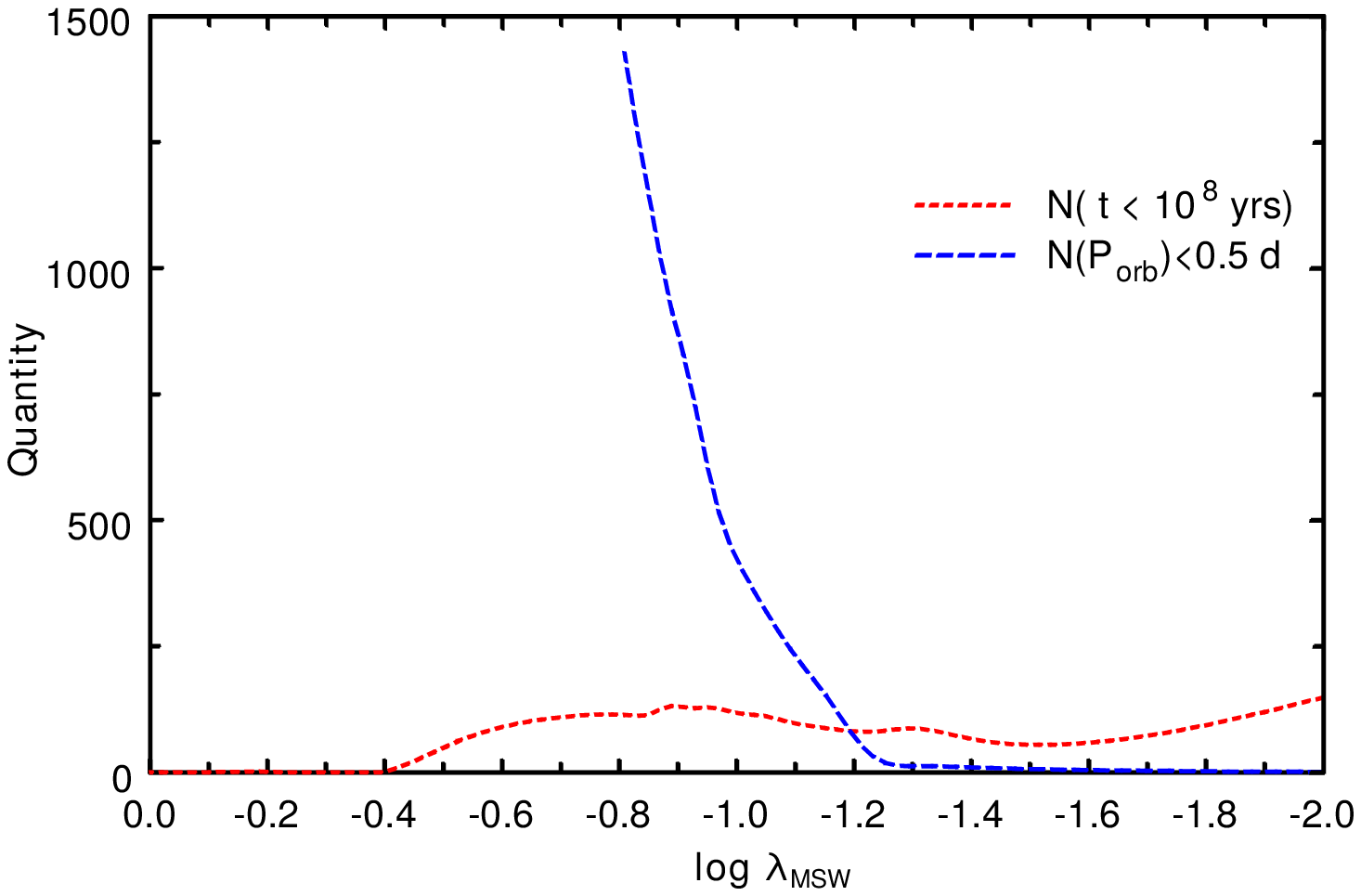}
\vspace{5pt} \caption{The quantity of X-ray novae in the Galaxy that costist of black holes and main sequence stars (including Roche lobe filling stars on the main sequence) with masses from $0.2R_{\odot}$ and $1.1R_{\odot}$ depending on the magnetic stellar wind strength. Curves show dependencies of the number of binaries with the orbital period less than 0.5 days and the number of systems with characteristic time scale of the angular momentum loss via the magnetic stellar wind less than 100 million years (e.g., all three mention X-ray novae). With the increase of the rate of the angular momentum loss with the magnetic stellar wind (i.e, with the decrease of $\lambda_{MSW}$) the quantity of binaries with the orbital period less than 0.5 days decreases, because the system in this range of orbital periods evolves faster and faster until the full ``sweeping'' of systems from this range. The quantity of systems with the characteristic time of the angular momentum loss via the magnetic stellar wind less than 100 million years is zero if the strength of the magnetic wind is usual, and it increases to a certain amount that weakly depends on the subsequent grow of the wind's strength, because more and more wide binaries correspond this definition. We choose the value $\lambda_{MSW}=0.13$, since in this case the ratio of the number of X-ray novae with measured masses of black holes to the number of X-ray novae with anomalously fast decrease of the orbital period approximately corresponds to observations (3 binaries with the fast decrease of the orbital period among 15 X-ray novae with known masses).}\label{f3}
\end{figure}

Fig. \ref{f3} shows two curves, one of them depicts the quantity of studied systems with orbital period less than 0.5 days, another curve depicts the quantity of systems with a characteristic time of the angular momentum loss via the magnetic stellar wind less than 100 million years. If the magnetic wind's strength is relatively low the second group contains no systems, with the increase of the strength the magnetic stellar wind time can become short even for systems with orbital periods longer than 0.5 days (see Fig. \ref{f2} which shows the quantity of binaries with $P_{orb}<1.5$), it can be seen that this quantity is enough to allow the stronger wind to bring wider pair closer keeping the number of close X-ray novae adequate to observations. Taking into account that the quantity of X-ray novae with anomalously short time scales of the orbital period decrease is about 15-20\% from all X-ray novae with measured masses of black holes it is possible to assume that $\lambda_{MSW}\approx 0.13$ is suitable to explain the observed quantity of X-ray novae. The chosen $\lambda_{MSW}$ corresponds to the increase of the angular momentum loss rate by the magnetic stellar wind by 60 times in comparison to the usual value.

Tables \ref{track1} and \ref{track2} shows examples of evolutionary tracks with the same initial parameters and the same evolutionary parameters except $\lambda_{MSW}$ in order to demonstrate the role of the strengthened magnetic stellar wind in the evolution. The evolution in both tracks is the same in both tables before the common envelope. Even very close approach of the stars after this stage produces no significant angular momentum loss by the magnetic stellar wind for $\lambda_{MSW}=1$ (Table \ref{track1}). In Table \ref{track2} ($\lambda_{MSW}=0.13$) after the common envelope the secondary (low mass) star becomes the source of the strong magnetic stellar wind that causes the strong angular momentum loss, therefore the semi-major axis grow due to the mass loss by the Wolf-Rayet star much slowly than in the previous case. It allows the Wolf-Rayet star to fill its Roche lobe, than the stars move closer to each other, and the secondary star also fills its Roche lobe, both stars lose significant parts of their masses. After the supernova explosion in case of the weak magnetic stellar wind the binary system consisting of the black hole and the low mass non-degenerate star remains, its life time if very long. In case of the strong magnetic wind the formal time of the angular momentum loss according to Equation \ref{msw-time} becomes very short, and the fate of the system becomes unclear. If the core of the non-degenerate star is not enriched with helium at that time when the mass value becomes equal to $0.3M_{\odot}$ the magnetic stellar wind can stop \citep{tutukov2018}, perhaps, the star can lose its magnetic stellar wind when its mass drops to the mass of a brown dwarf or even a giant planet \citep{paczynski1981}, however the possibility of the unlimited approach of the star to the black hole until its disintegration also remains.

\subsection{Conclusions from evolutionary calculations}

In highly magnetized active stars the angular momentum is able to re-distribute, and the star can deform. The deformation can lead to the change of the quadrupole gravitational momentum of the system \citep{applegate1992}. This process can cause quasi-periodical variations of the orbital period up to $10^{-5}$ in time scales up to several decades. This mechanism potentially can be involved to explain the rapid approach of components of studied systems. Nevertheless, taking into account that all three systems with the anomalous change of the orbital period show only the decrease of the period, this is rather not the main explanation.

The existence of close pairs consisting of a black hole and a very low mass non-degenerate star means that more than one half of the pre-supernova mass collapses to the black hole (otherwise the system unavoidably decays). Low or zero natal kick of the black hole during its formation is also required. Systems A0620-00 and Nova Muscae 1991 belong to the galactic disk confirming this statement. KV UMa is at the distance more than 1 kpc from the galactic plane, that potentially can be an evidence in favour of a significant natal kick for this system. If so, to keep the binarity of KV UMa the kick should have a very specific direction and a limited value (since a strong kick can disrupt the system independently on its direction). Probably, KV UMa got its additional velocity not in the supernova explosion process, but in dynamical interactions with other bodies.

The black hole paired with the normal star with the orbital period about several hours most likely was formed due to the common envelope stage, because the semi-major axis of the system is less (or very close) to the radius of the main sequence progenitor of the black hole. In case of X-ray novae the common envelope could be a source of strong magnetic fields in the non-degenerate star. So, the angular momentum loss by the strengthened magnetic stellar wind as the explanation of the anomalous braking in studied systems has a physical basis. To explain the observed composition and quantity of X-ray novae it should be assumed the increase of the angular momentum loss via the magnetic stellar wind by 60 times in comparison to the usually used value. The value $\lambda_{MSW}=0.13$ can be a bit different if a different description is used, as, for example, in papers by \citet{verbunt1981,vilhu1982,romani1992,romani1994,pz1997,podsiadlowski2002,yungelson2006}, but the main conclusion of this evolutionary study does not change: the magnetic field of the low mass non-degenerate star can be increased during the common envelope stage leading to the increased magnetic stellar wind in further evolution of the system. Nevertheless, potentially the observed rapid approach of the components of studied binaries can have an alternative explanation, e.g., using a circumbinary disk around both components \citep{chen2015,xu2018}.




\bsp	
\label{lastpage}
\end{document}